\documentclass[fleqn,usenatbib]{mnras}

\usepackage{graphicx}
\usepackage{ifthen}
\usepackage{url}
\usepackage{amsmath}
\usepackage{bm}
\usepackage{lscape}
\usepackage{rotating}
\usepackage[graphicx]{realboxes}
\usepackage{supertabular}
\usepackage{pdfpages}
\usepackage{footnote}
\usepackage{hyperref}
\usepackage{amssymb}
\makesavenoteenv{tabular}
\makesavenoteenv{table}
\usepackage[T1]{fontenc}
\usepackage{tipa}
\usepackage{accents}
\newcommand{\ubar}[1]{\underaccent{\bar}{#1}}
\usepackage{tablefootnote}

\hypersetup{
  colorlinks   = true,    
  urlcolor     = blue,    
  linkcolor    = blue,    
  citecolor    = blue      
}

\usepackage{ulem}

\def\ltsima{$\; \buildrel < \over \sim \;$}
\def\lta{\lower.5ex\hbox{\ltsima}}
\def\gtsima{$\; \buildrel > \over \sim \;$}
\def\simgt{\lower.5ex\hbox{\gtsima}}
\def\kms{{\rm\,km \  s^{-1}}}

\def\msun{{\rm\,M_\odot}}

\usepackage{amsmath}

\def\s{\ifmmode \widetilde \else \~\fi}
\def\={\overline}

\def\spose#1{\hbox to 0pt{#1\hss}}

\def\eg{{e.g.,\ }}
\def\ie{{i.e.,\ }}
\def\lta{\mathrel{\spose{\lower 3pt\hbox{$\mathchar"218$}}
     \raise 2.0pt\hbox{$\mathchar"13C$}}}
\def\gta{\mathrel{\spose{\lower 3pt\hbox{$\mathchar"218$}}
     \raise 2.0pt\hbox{$\mathchar"13E$}}}
\def\Dt{\spose{\raise 1.5ex\hbox{\hskip3pt$\mathchar"201$}}}    
\def\dt{\spose{\raise 1.0ex\hbox{\hskip2pt$\mathchar"201$}}}    

\def\dotsfill{\leaders\hbox to 1em{\hss.\hss}\hfill}

\def\FeH{{\rm[Fe/H]}}

\def\ii{{~\sc ii}}

\loadboldmathitalic 
\title[Stars on the edge: Sculptor]{Stars on the edge: Galactic tides and the outskirts of the Sculptor dwarf spheroidal} 
\author[F. Sestito et al.] {Federico Sestito$^{1}$\thanks{Email: \url{sestitof@uvic.ca}},
Joel Roediger$^{2}$,
Julio F. Navarro$^{1}$,
Jaclyn Jensen$^{1}$,
Kim A. Venn$^{1}$,
\newauthor 
Simon E. T. Smith$^{1}$,
Christian Hayes$^{2}$, and
Alan W. McConnachie$^{2,1}$
\\
$^{1}$ Department of Physics and Astronomy, University of Victoria, PO Box 3055, STN CSC, Victoria BC V8W 3P6, Canada\\
$^{2}$ NRC Herzberg Astronomy \& Astrophysics, 5071 West Saanich Road, Victoria, BC V9E 2E7, Canada\\
}

\pubyear{2023}
\date{Accepted XXX. Received YYY; in original form ZZZ}
\begin{document}
\maketitle 
\label{firstpage}
\pagerange{\pageref{firstpage}--\pageref{lastpage}}

\begin{abstract}
The formation of "stellar halos" in dwarf galaxies have been discussed in terms of early mergers or Galactic tides, although fluctuations in the gravitational potential due to stellar feedback is also a possible candidate mechanism. A Bayesian algorithm is used to find new candidate members in the extreme outskirts of the Sculptor dwarf galaxy. Precise metallicities and radial velocities for two distant stars are measured from their spectra taken with the Gemini South GMOS spectrograph. The radial velocity, proper motion and metallicity of these targets are consistent with Sculptor membership. As a result, the known boundary of the Sculptor dwarf extends now out to an elliptical distance of $\sim10$ half-light radii, which corresponds to a projected physical distance of $\sim3$ kpc. As reported in earlier work, the overall distribution of radial velocities and metallicities indicate  the presence of a more spatially and kinematically dispersed metal-poor population that surrounds the more concentrated and colder metal-rich stars. Sculptor's density profile shows a "kink" in its logarithmic slope at a projected distance of $\sim25$ arcmin (620 pc), which we interpret as evidence that Galactic tides have helped to populate the distant outskirts of the dwarf. We discuss further ways to test and validate this tidal interpretation for the origin of these distant stars.
\end{abstract}


\section{Introduction}
The dwarf satellites orbiting the Milky Way are among the oldest and most metal-poor galaxies that are known to exist \citep[e.g.,][]{Tolstoy09}. These galaxies have a broad range of masses, sizes, and luminosities. A luminosity threshold of L$= 10^5$ L$_{\odot}$ was proposed by \citet{Simon19} to differentiate between classical dwarf galaxies (such as Fornax, Sculptor, and Ursa Minor) and the least massive ultra-faint dwarf galaxies \citep[UFDs, also see][]{Bullock17}. The UFDs are situated at the low-mass end of the hierarchical formation process, with just enough mass to produce very metal-poor stars \citep[VMP, with $\FeH{} \leq-2.0$,][]{Simon19}. The dwarf satellites are a prime target to comprehend the characteristics of dark matter and the early stages of our galaxy's formation \citep[e.g.,][]{Bullock17}.

The presence of extended stellar halos around dSphs is of particular interest. In the context of the hierarchical formation of $\Lambda-$Cold Dark Matter ($\Lambda-$CDM) cosmology \citep[\eg][]{White78,Frenk88,NavarroFrenkWhite97}, stellar haloes are built primarily through the accretion and disruption of smaller systems. While this process is clearly at work in large galaxies like the Milky Way (MW), the possible presence of stellar haloes around dSphs  remains inconclusive \citep[\eg][and references therein]{Deason22}. In $\Lambda-$CDM, the fraction of stars accreted through mergers is thought to become negligible in the regime of dwarfs, suggesting, at face value, that dwarfs should generally lack stellar halos \citep[\eg][]{Genel10,Moster13}. 

Faint dwarf galaxies have shallow gravitational potentials, and are therefore extremely susceptible to a number of processes that  may act to influence their individual morphologies and perhaps lead to the formation of an extended envelope of stars \citep[\eg][and references therein]{Higgs21}.  Internal processes include star formation and the subsequent stellar feedback \citep[\eg][]{ElBadry18b};  external processes include mergers \citep[\eg][]{Deason14}, ram pressure stripping \citep[\eg][]{Grebel03} and stirring \citep[\eg][]{Kazantzidis11}, tidal interaction \citep[\eg][]{Fattahi18}, and reionization \citep[\eg][]{Wheeler19}. The outskirts of a dwarf satellite galaxy is fertile ground to look for signatures of these internal and external processes.

Very recently, new candidate member stars have been reported in the extreme outskirts of several dwarf galaxies \citep[\eg][]{McVenn2020a,Chiti21,Filion21,Longeard22,Yang22,Jensen23,Sestito23Umi,Waller23}. \citet{Chiti21} spectroscopically identified members as far out as $\sim$9 half-light radii ($r_h$), or physical distances up to 1 kpc, away from the centre of Tucana~II. Their chemo-dynamical investigation showed that the outer envelope of Tucana II is more metal-poor than the central regions, a result that suggests that Tucana~II was probably formed by an early merger which dispersed the old stellar component and brought gas to the inner regions to fuel the formation of subsequent generations of stars \citep{BenitezLlambay16}.

Recently, \citet{Jensen23} improved a Bayesian algorithm,  developed by \citet{McVenn2020a}, to estimate the  probability that a star in the vicinity of a dwarf galaxy is a member of the dwarf. Their method combines the full astrometric and photometric data from \textit{Gaia} Early Data Release 3 \citep[EDR3,][]{GaiaEDR3}. The algorithm is able to identify as members the distant stars found by \citet{Chiti21} around Tucana~II \citep{Jensen23}, and it has been used to hunt for an extended stellar halo in Coma Berenices, Ursa Major~I, Bo{\"o}tes~I  \citep[$\sim6\ \ r_h$,][]{Waller23}, and Ursa Minor \citep[$\sim12\ \ r_h$,][]{Sestito23Umi}. Both \citet{Waller23} and \citet{Sestito23Umi} discussed that the outermost stars in these systems may have formed in the central regions, then moved out through tidal stripping and/or supernova feedback.

Other algorithms have also been applied to search for extended stellar haloes in  faint systems. \citet{Yang22} report a new halo component in Fornax, representing about 10 percent of the total stellar mass; \citet{Longeard22} combined Gaia EDR3 data with photometric metallicites from the Pristine survey \citep{Starkenburg17b} to discover 27 new members in the  outskirt of Bo{\"o}tes I. In the same system, \citet{Filion21} found several blue horizontal branch stars beyond its tidal radius.

In this work, candidates identified by this extremely efficient Bayesian algorithm \citep{Jensen23}  are used to confirm the presence of stars in the outermost region of the Sculptor dwarf galaxy (Scl).  The two outermost candidate member stars at $\sim10$ half-light radii are observed with low-resolution spectroscopy.  Their radial velocities and metallicities confirm their membership to Sculptor.

Sculptor is the first faint MW satellite discovered \citep{Shapley38} after the Magellanic Clouds.  Since then, the number of known faint dwarf galaxies has grown tremendously thanks to new observational techniques and survey, \eg the Sloan Digital Sky Survey \citep[SDSS,][]{York00}, the Panoramic Survey Telescope And Rapid Response System \citep[Pan-STARRS,][]{Chambers16}, the Dark Energy Survey \citep[DES,][]{DES05}, the DESI Legacy Imaging Surveys \citep{DESI19}, the Pan-Andromeda Archaeological Survey \citep[PAndAS, \eg][]{Martin09,McConnachie09}, and the DECam Local Volume Exploration Survey \citep[DELVE,][]{DELVE21}. Many spectroscopic campaigns have targeted stars in Sculptor. In particular, the Dwarf Abundance and Radial velocity Team  \citep[DART,][]{Tolstoy06,Battaglia08b} observed more than 600 members measuring precise metallicities and radial velocities, showing that the system spans 3 dex in metallicity.

High-resolution spectroscopic observations probing the chemical properties of Sculptor are more scarce. There are only 12 stars with $\FeH<-2.5$ observed so far \citep{Frebel10, Tafelmeyer10, Starkenburg13, Jablonka15, Simon15,Skuladottir21}. The first ultra-metal-poor star (UMP, $\FeH\leq-4.0$) observed at high-resolution in a dwarf galaxy is a Sculptor member, as recently discovered by \citet{Skuladottir21} \citep[vs. 42 detected UMPs in the Milky Way,][]{Sestito19}. The authors discuss that the atmosphere of this ancient star carries the imprints of a hypernova generated from a metal-free star, also known as First star or Population III. This finding is in agreement with \citet{Ishigaki18}, in which they discuss that the the majority of the primordial supernovae would have masses $<40\msun$. However, \citet{Hartwig23} showed that only $\sim40$ percent of the stars at that metallicity \citep[$\FeH=-4.11$,][]{Skuladottir21} would be mono-enriched by primordial supernovae. The chemical evolution of Sculptor is similar to other dwarf galaxies. At  very low-metallicities, Sculptor is dominated by contributions from supernovae type II \citep[\eg][]{Jablonka15,Skuladottir21}, while at $\FeH\geq-1.9$ indications of supernovae type Ia are observable in the distribution of  chemical abundances \citep[\eg][]{Venn04,Tolstoy09,Hill19}. Star formation in Sculptor has been modelled to have lasted between 6 to 8 Gyr \citep[\eg][]{Fenner06,Revaz18}. 

This paper is divided as follow. Section~\ref{sec:data} reports the target selection, the observational setup, and the spectra reduction of the two targets. Section~\ref{sec:stellarparam} describes the inference of the stellar parameters. Section~\ref{Sec:rvmet} outlines the measurement of the radial velocity and the metallicity from the near infrared Ca Triplet lines. The discussion on the membership of the targets and the presence of tidal perturbations in Sculptor are reported in Section~\ref{sec:discussion}. Conclusions are summarised in Section~\ref{sec:conclusions}.

\section{Data}\label{sec:data}

\subsection{Target selection}
Candidate member stars for spectroscopic follow-up are firstly selected using the algorithm described in \citet[][hereafter J23]{Jensen23}. The algorithm, much like its predecessor described in \citet[][hereafter MV20]{McVenn2020a}, is designed to search for member stars in a given dwarf galaxy by determining the probability of membership to the satellite. The probability of being a satellite member, P$_{\rm{sat}}$, is determined using three likelihoods based on the dwarf’s (1) colour-magnitude diagram (CMD), (2) systemic proper motion, and (3) the radial distance from the center of the satellite, using relevant data from \textit{Gaia} EDR3 \citep[][]{GaiaEDR3}.  The stellar density of a dwarf can often sufficiently be summarized by a single exponential profile (see MV20). However, motivated to search for dwarfs which may have evidence of tidal features or extended stellar haloes, the spatial likelihood in J23 assumes that each dwarf may host a secondary, extended, and lower density, outer profile. Only a handful of systems are found to host a secondary outer profile (see J23), a few of which are already known to be actively tidally disrupting (\eg Bootes III and Tucana III). Sculptor is a system for which a secondary outer profile is observed, indicating either an extended stellar halo or tidal features. 

The most radially distant (9.7 and 10.1 $r_h$), relatively bright (G $\sim 17.6-17.9$ mag) candidates of Sculptor are selected.  J23 demonstrated that the purity of candidate members is extremely high for P$_{\rm{sat}}>0.40$ and the majority of  foreground stars has P$_{\rm{sat}}<0.10$. Therefore, the algorithm is excellent at removing foreground contaminants even at such low probabilities.  As a matter of fact, these stars are also listed as  likely Scl members by \citet[][]{Qi22}, with a probability $>98$ percent. Thus far, this algorithm has been proved to be very efficient in finding new members in the extreme outskirts of Coma Berenices, Ursa Major 1, Bootes 1 \citep{Waller23}, and in Ursa Minor \citep{Sestito23Umi}.

For the most distant candidate, Target~2, a measurement of its radial velocity is already present in the literature. \citet{Westfall06} measured radial velocities of stars in the direction of Scl. They found a radial velocity (RV) compatible with Scl ($\sim119\kms$). Given its large distance from Scl centre, the authors had no reason to suspect Target~2 was a member, therefore, no metallicity was estimated.
The main properties of Sculptor and our two targets are reported in Tables~\ref{tab:scuprop} and \ref{tab:targets}. 

\begin{figure*}
\includegraphics[width=1\textwidth]{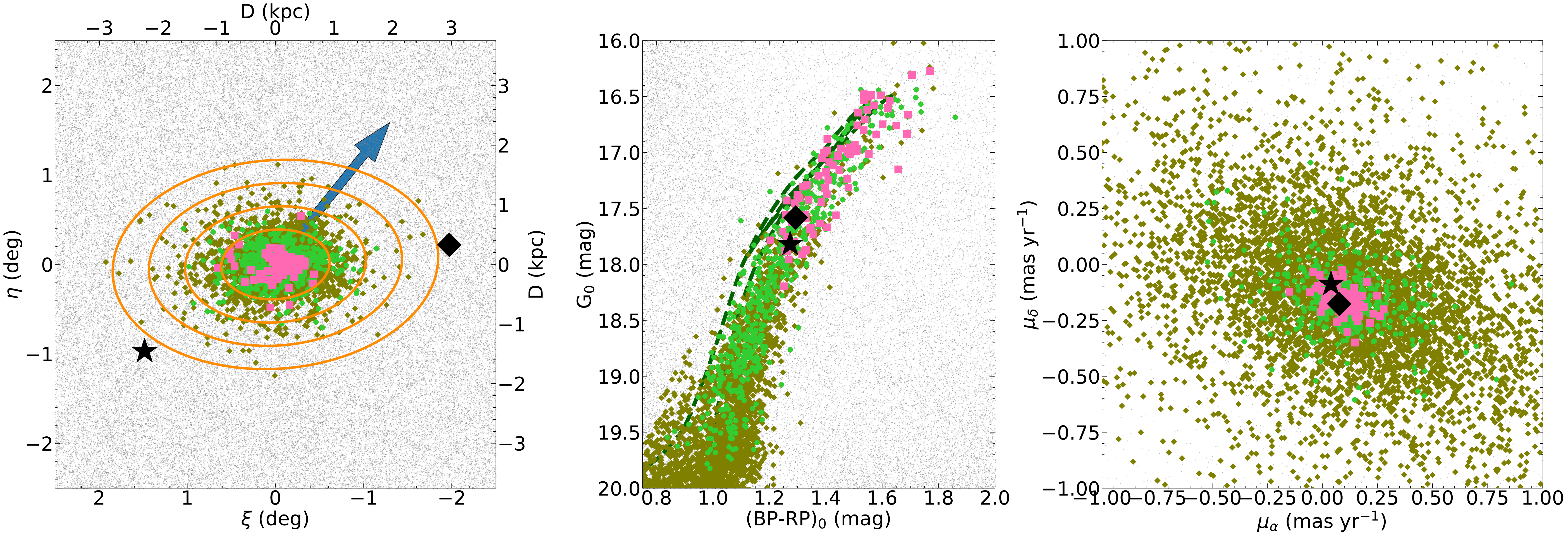}
\caption{Sculptor as seen in \textit{Gaia} EDR3. All  panels: Targets~1~and~2 are marked with a diamond and a star marker, respectively. Olive diamonds are candidate members from \citet{Jensen23}. Lime green circles and hot pink squares are Sculptor stars from the cleaned DART sample and APOGEE DR17, respectively. MW foreground stars are shown with small grey  dots. These are selected from \textit{Gaia} EDR3 in the direction of Sculptor and within the field of view of the $\eta-\xi$ panel. Left panel: Projected sky coordinates and projected distance from Sculptor's centre. The orange ellipses denote  elliptical distances of 3, 5, 7, and 9 $r_h$. The arrow points in the direction of Sculptor's proper motion. Central panel: Colour-magnitude diagram. Dark green dashed line is a Padova isochrone at $\FeH = -2.0$ and age of 12 Gyr \citep{Bressan12}. Right panel: Proper motion space.}
\label{Fig:onsky}
\end{figure*}

Figure~\ref{Fig:onsky} shows the position of our two targets (black diamond and star markers), $\sim7000$ candidate members selected from J23 (olive diamonds), and other known Sculptor members (lime green and hot pink markers) in  projected sky coordinates, on  colour-magnitude diagram, and in  proper motion space. The selection algorithm is also applied to stars within the Dwarf Abundance and Radial velocity Team  \citep[DART,][]{Tolstoy06,Battaglia08b} and to the APOGEE data release 17 \citep[DR17,][]{APOGEEDR17}. From the initial sample of DART (656 stars), stars with low probability membership ($<40$ percent, 118 stars) have been removed. Then, for the cleaned DART sample in common with  APOGEE DR17,  radial velocities and metallicities from the former are used, since the metallicity grid in APOGEE stops at \FeH{}$=-2.4$ and several stars have  metallicity below that threshold according to DART. We also note that RV in APOGEE DR17 are in agreement with DART even at $\FeH < -2.4$. The total sample with reliable radial velocities and metallicities is composed of 617 stars, 538 from DART and 79 from APOGEE DR17.

\begin{table}
\caption[]{Galactic parameters of Sculptor. The coordinates $\alpha,\delta$, the mean metallicity, the  mean radial velocity, the dispersion in velocity, the heliocentric distance  D$_\odot$, the distance modulus, the ellipticity, the position angle $\phi$, and the  half-light radius  r$_{\rm h}$ in arcmin and pc, the Gaia EDR3 mean proper motion, and the dynamical mass  are reported with the respective references. (a) refers to \citet{Mcconnachie12}, (b) to \citet{McVenn2020a}, and (c) to \citet{McVenn2020b}.}
\centering
\resizebox{0.47\textwidth}{!}{
\hspace{-0.6cm}
\begin{tabular}{lrc}
\hline
Property & Value & Reference\\
\hline
 $\alpha$ &15.03917 deg  &  (b) \\
 $\delta$ & $-33.70917$ deg & (b) \\
 $\overline{\FeH}$ & $-1.68 \pm0.01$ & (b) \\
  $\overline{\rm{RV}}$ & $111.4$ km s$^{-1}$ & (a) \\
  $\sigma_v$ & $9.2\pm 1.4$ km s$^{-1}$ & (b)\\
 D$_\odot$   & $86\pm6$ kpc & (a) \\
 D$_{\rm{mod}}$ & $19.67\pm0.14$ mag & (a)\\
ellipticity  &  $0.37\pm0.01$ & (b)\\
 $\phi$   &  $94\pm1 $ deg & (b) \\
  r$_{\rm h}$ & $12.33\pm 0.05$ arcmin  & (b)\\
  r$_{\rm h}$ & $308\pm 20$ pc & (b) \\ 
  $\mu_{\alpha}\rm{cos}\delta$ & $0.099 \pm 0.002$ mas yr$^{-1}$  & (c)\\
   $\mu_{\delta}$ & $-0.160 \pm 0.002$ mas yr$^{-1}$  & (c) \\
    M$_{\rm{dyn}}$($\leq$r$_{\rm{h}}$) & $14\times10^6\msun$& (a)\\
\hline \hline
\end{tabular}}
\label{tab:scuprop}
\end{table}

\begin{table*}
\caption[]{The \textit{Gaia} EDR3 source ID, the coordinates $(\alpha,\delta)$, the projected coordinates $(\xi,\eta)$, the elliptical radius distance $r_{\rm{ell}}$ in $r_h$ unit, the probability to be a member from \citet{Jensen23}, the \textit{Gaia} EDR3 photometry G and BP$-$RP, and the reddening A$_{\rm V}$ from \citet{Schlafly11} are reported for each target.}
\centering
\resizebox{1\textwidth}{!}{
\hspace{-0.6cm}
\begin{tabular}{lcccccccccc}
\hline
Target &  source id & $\alpha$ & $\delta$ & $\xi$  & $\eta$  & $r_{\rm{ell}}$ & $P_{\rm{sat}}$ & G & BP$-$RP & A$_{\rm V}$ \\
 & & (deg)& (deg) & (deg) & (deg) & ($r_h$) & & (mag) & (mag)& (mag)   \\ \hline
 Target~1  & 5006419626331394048 & $12.67539$ & $-33.46838$ & $-1.97246$ & $0.21835$ & $9.67$ &  $0.46$ & $17.62$ &  $1.31$ &  0.0419  \\
Target~2  & 5026130884816022016 & $16.84351$ & $-34.66460$ & $1.48453$ & $-0.96882$ & $10.06$ &  $0.45$ & $17.86$ &  $1.29$ &  0.0439 \\
\hline
\end{tabular}}
\label{tab:targets}
\end{table*}

\subsection{GMOS observations and reduction}
Targets are observed with the Gemini Multi-Object Spectrograph \citep[GMOS,][]{Hook04,Gimeno16} at the Gemini South with a Fast Turnaround program, GS-2022B-FT-205 (P.I. F. Sestito). The R831 grating centred at the 855 nm Ca\ii{} Triplet line, with a slit width of 1.0 arcsec, is used for these observations. Each target is observed with 4 exposures of 900s each to obtain a signal-to-noise ratio per spectral pixel of $\sim$50 in the Ca\ii{} Triplet region, in order to measure the RV with an uncertainty around $\sim10$ $\kms$ and the equivalent width (EW) of the Ca\ii{} Triplet with a sufficient precision for reaching an uncertainty of $\sim0.3$ dex in the metallicity estimate. The latter is inferred through various  Ca\ii{} T EW~-~\FeH{} relations, as described in Section~\ref{Sec:rvmet}.

Spectra have been reduced with the standard routines for GMOS data in the Gemini \textsc{IRAF} package\footnote{\url{http://www.gemini.edu/observing/phase-iii/understanding-and-processing-data/data-processing-software/gemini-iraf-general}} \citep{Tody86,Tody93}. Spectra have been corrected for bias and flat-fields, and the wavelength has been calibrated  using the CuAr emission lamp. Heliocentric motion has been corrected in the observations. The 4 exposures have been stacked using \textsc{GemCombine}. The flux of the combined spectra have been normalised to one. The log of the observations is reported in Table~\ref{tab:obs}.

\begin{figure}
\includegraphics[width=0.47\textwidth]{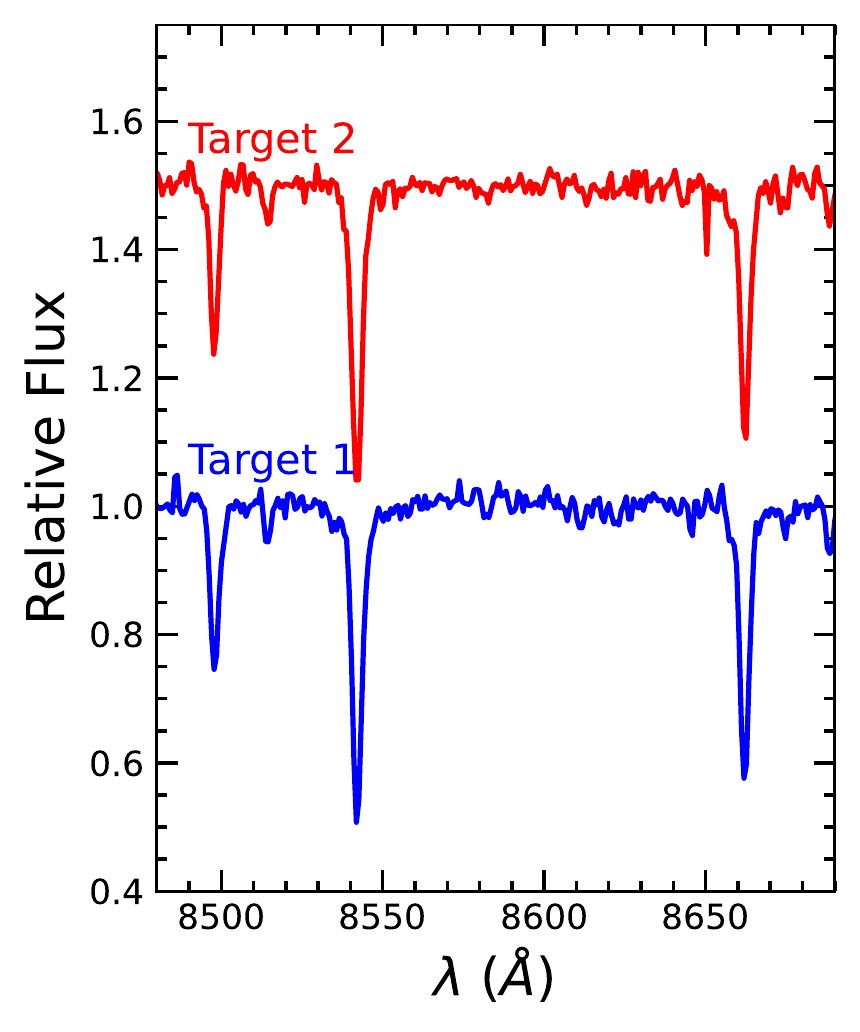}
\caption{Normalised GMOS spectra in the Ca T region.}
\label{Fig:spectra}
\end{figure}

Figure~\ref{Fig:spectra} displays the the Ca\ii{} region of the combined and normalised GMOS spectra of Target~1~and~2.

\begin{table}
\caption[]{Total exposure time, number of exposures,  signal-to-noise ratio (SNR) measured at the Ca\ii{} T region, and the observation dates are reported for each target. The SNR is calculated as the ratio between the median flux and its standard deviation in the region [8550,8650] \AA.
}

\resizebox{0.48\textwidth}{!}{
\hspace{-0.6cm}
\begin{tabular}{lcccc}
\hline
Target  & $t_{\rm exp}$ & N$_{\rm exp}$  & SNR & Obs. date\\ 
  & (s) &   & @Ca\ii{} T & YY/MM/DD \\ \hline
Target~1  & 3600 & 4 & 66 & 22/11/15 \\
Target~2  &3600 &4 & 78 &  22/10/30 \\
\hline
\end{tabular}}
\label{tab:obs}
\end{table}

\section{Stellar parameters}\label{sec:stellarparam}
To estimate the effective temperatures, the \citet[][]{Mucciarelli21} colour-temperature relationship for giants is used in this work. This is based on the Infrared flux method from \citet{Gonzalez09} and adapted to {\it Gaia} EDR3 photometry. The input parameters are the {\it Gaia} EDR3 (BP-RP) de-reddened colour and a metallicity estimate. The 2D \citet{Schlafly11} map\footnote{\url{https://irsa.ipac.caltech.edu/applications/DUST/}} has been used to correct the photometry for extinction\footnote{To convert from the  E(B-V) map to  {\it Gaia} extinction coefficients,  the  $\rm A_V/E(B-V)= 3.1$ \citep{Schultz75} and the $\rm A_G/A_V = 0.85926$, $\rm A_{BP} /A_V = 1.06794$, $\rm A_{RP} /A_V = 0.65199$ relations \citep{Marigo08,Evans18} are used.}.  A value of $\FeH=-2.0\pm0.5$ is adopted as input metallicity, compatible with the distribution of \FeH{} in Sculptor.

Surface gravities were inferred from the Stefan-Boltzmann equation\footnote{$L_{\star} = 4\pi R_{\star}^2 \sigma T_{\star}^4$; the radius of the star can be calculated from this equation, then the surface gravity is inferred assuming the mass.}. This calculation requires the effective temperature, the distance of the object (see Table~\ref{tab:scuprop}), the {\it Gaia} EDR3 G de-reddened photometry, and the bolometric corrections on the flux \citep{Andrae18}.  A Monte Carlo algorithm has been applied to the input parameters with their uncertainties to estimate the uncertainties on the stellar parameters. The input quantities were drawn  from a  Gaussian distribution, except for the stellar mass.  The latter is treated with a flat prior from 0.5 to 0.8 $\msun$, which is consistent with the mass of  very metal-poor stars. The mean uncertainty on the effective temperature is $\sim 72$~K, while on the surface gravity  is $\sim 0.08$~dex, both mainly driven by the distance's uncertainty.

Stellar parameters from these methods have been shown to be in agreement with spectroscopic methods applied to very metal-poor stars \citep[\eg][]{Kielty21,Lardo21,Sestito23,Waller23}. 
The  stellar parameters are reported in Table~\ref{tab:params}.

\begin{table}
\caption[]{Effective temperatures and surface gravities of the targets.}
\centering
\begin{tabular}{lcc}
\hline
Target  & T$_{\rm{eff}}$ & log~g  \\ 
 &  (K)&   \\ \hline 
 Target~1  &   $4535 \pm 75$ & $1.11 \pm 0.07$  \\
Target~2   & $4583 \pm 70$ & $1.24 \pm 0.08$ \\ 
\hline
\end{tabular}
\label{tab:params}
\end{table}

\section{Metallicity and radial velocity estimation}\label{Sec:rvmet}

\subsection{Radial velocity measurement}
The radial velocity has been calculated by cross-correlating the combined normalised spectra with a synthetic spectrum using the \textsc{fxcor} routine in \textsc{IRAF}. The synthetic spectrum has been created from \textsc{synth} \textsc{MOOG} \citep{Sneden73} using a spectral line list from \textsc{linemake}\footnote{\url{https://github.com/vmplacco/linemake}} \citep{Placco21}. \textsc{MARCS} model atmospheres \citep{Gustafsson08,Plez12} are adopted with the same stellar parameters as the observed stars. The synthetic spectrum has been created at the same resolution of the GMOS R831 grating.

To test for possible RV offsets due to problems in the wavelength solution, telluric absorption lines have been extracted from the spectra of our targets and cross-correlated. There is no RV offset between the spectra of the two targets within the dispersion of GMOS spectrograph ($\sim13\kms$).

\subsection{Metallicities from Ca\ii{} Triplet}
Multiple calibrations to measure \FeH{} from the near-infrared Ca\ii{} Triplet lines ($\lambda\lambda 8498,8542,8662$ \AA) are available in the literature. In this work, calibrations from  \citet[][hereafter S10]{Starkenburg10}, \citet[][hereafter C13]{Carrera13}, and \citet[][hereafter L23]{Longeard23} are selected. All of them are in the form: 
\begin{equation}
    \FeH = a+b\cdot \rm{M_V} + c\cdot \rm{\Sigma Ca} + d\cdot \rm{\Sigma Ca}^{-1.5} + e\cdot \rm{\Sigma Ca} \cdot \rm{M_V},
    \label{Eq:ewmet}
\end{equation}
where $a,b,c,d$ are coefficients, $\rm{M_V}$ is the absolute V magnitude of the giant star, $\rm{\Sigma Ca}$ is a linear function of the EWs of the components of the Ca\ii{} T lines. In the case of the calibration from S10,  $\rm{\Sigma Ca}$ is the sum of the EWs of the second and third components of the Triplet, $\rm{\Sigma Ca} = \rm{EW}_{2} +\rm{EW}_{3}$. The calibration from C13 uses $\rm{\Sigma Ca}$ as the weighted sum of the three components,  $\rm{\Sigma Ca} = 0.19\cdot\rm{EW}_{1} +0.47\cdot\rm{EW}_{2} +0.34\cdot\rm{EW}_{3} $. L23 use the first and the second lines of the triplet, $\rm{\Sigma Ca} = \rm{EW}_{1} +\rm{EW}_{2}$.

$\rm{M_V}$ is derived converting the \textit{Gaia} EDR3 magnitudes to the Johnson-Cousin filter following \citet{Riello21} and adopting a distance modulus of $19.67\pm0.14$ mag (see Table~\ref{tab:scuprop}). The EW is measured using the \textsc{splot} routine in \textsc{IRAF}, fitting the spectral lines with multiple profiles. The median and the standard deviation of the multiple measurements have been adopted as values for the EW and its uncertainty. A Monte Carlo with $10^6$  randomisations is performed on the input variables of Equation~\ref{Eq:ewmet} and on the coefficients drawing them from a Gaussian distribution. In case of the coefficients from L23, the randomisations have been done taking into account their strong correlations (see L23). While the relative uncertainty on the coefficients from S10 and C13 is set to 10 percent. The \FeH{} and its uncertainty are the median and the standard deviation of the draws, respectively. Metallicities and radial velocities are reported in Table~\ref{tab:met}. The mean of the three metallicity estimates is reported as well. We note that the metallicities from S10 and C13 differ by $\sim1.35\sigma$, while the ones from S10 and L23 by only $\sim0.33\sigma$. One source of discrepancy might be the different weight in $\rm{\Sigma Ca}$ given to the Ca\ii{} T lines by the different calibrations. While they have the same weights in S10 and L23, the first component of the triplet has a less impact in C13 (first : second is $1:2.47$ and first : third is $1:1.79$). As discussed below, the small discrepancy in the metallicities does not affect the membership of the two candidates.

Figure~\ref{Fig:rv_met_rh} displays the metallicities and radial velocities of our targets and known Sculptor members (from DART and APOGEE DR17) as a function of their elliptical distances (left panels); the \FeH{} vs. RV space and their histograms (central and right panels). The three calibrations provide \FeH{} for both Target~1~and~2 in agreement with the Sculptor's range.  As regards Target~1, we note that our RV determination places this star at the edge of the distribution for this system. Its RV is compatible with the value from a few known members observed by the DART survey and located in the inner region ($\lesssim6r_h$) of the system (see top left panel of Figure~\ref{Fig:rv_met_rh}). The  RV of Target~2 is in agreement with the Sculptor distribution  and also with the previous determination from \citet{Westfall06}.

\subsection{Comparison with the Besan\c{c}on model}
\begin{figure*}
\includegraphics[width=\textwidth]{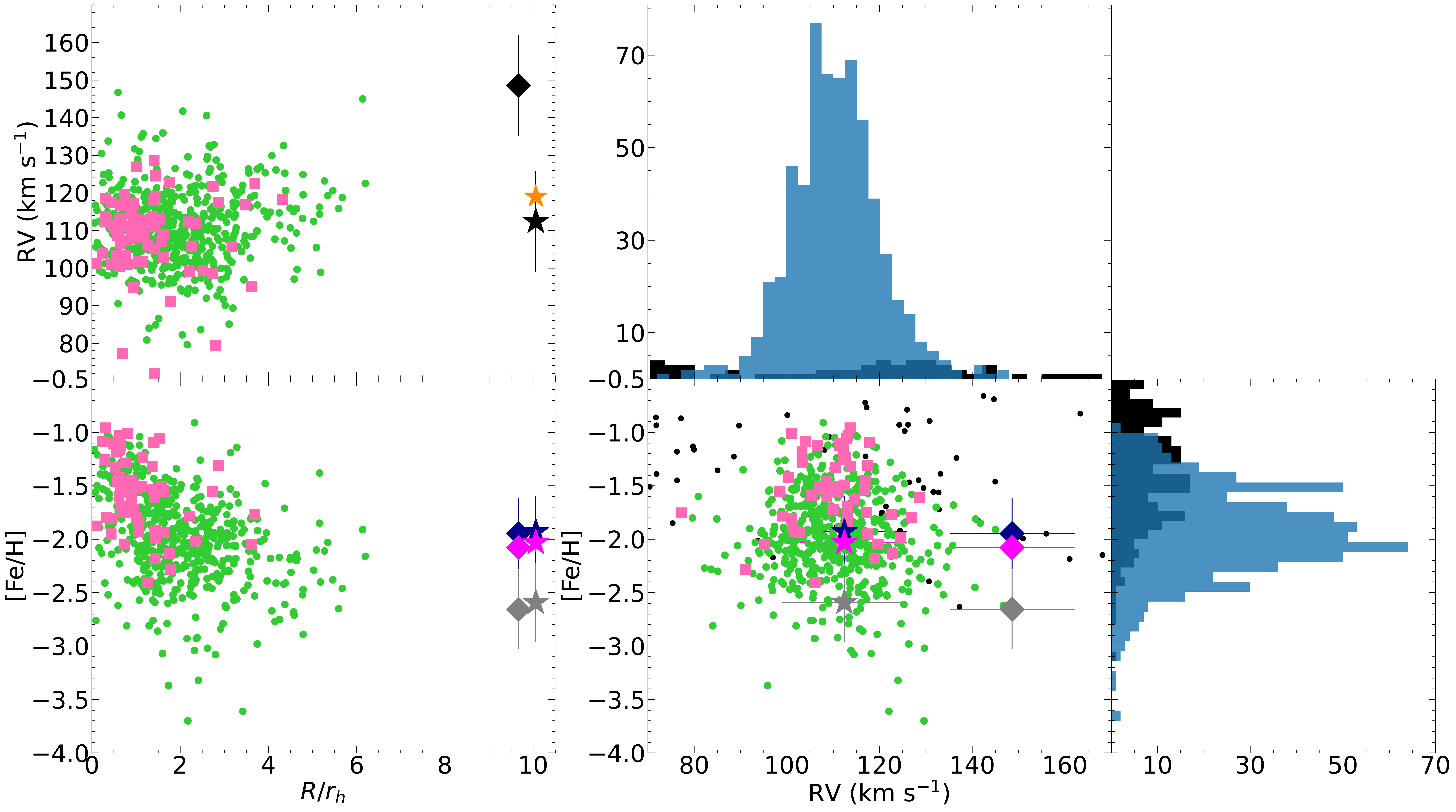}
\caption{Distributions of the Sculptor stars. Left panels: Radial velocities (top) and metallicities (bottom) as a function of the elliptical distance. Central panel: distribution of Sculptor stars in the \FeH{} vs. RV space. Corner plots: histograms of the RV (top) and metallicity (right) distributions. Targets~1~and~2 are marked with a diamond and a star marker, respectively. The metallicities from three calibration are denoted by dark blue \citep{Starkenburg10}, grey \citep{Carrera13}, and magenta \citep{Longeard23} markers. RV of Target~2 from \citet{Westfall06} is denoted by a dark orange star, which is in agreement with our measurement. Lime green dots are the cleaned sample from DART, while hot pink squares are  Sculptor members from APOGEE DR17. Stellar particles from the Besan\c{c}on simulation are marked with black in the RV vs. \FeH{} and the histograms panels.}
\label{Fig:rv_met_rh}
\end{figure*}

Figure~\ref{Fig:rv_met_rh} also shows the  Besan\c{c}on simulation\footnote{\url{https://model.obs-besancon.fr}} \citep{Robin03,Robin17} of the Milky Way halo towards Sculptor, with the same cut on sky position as in the left panel of Figure~\ref{Fig:onsky}. Nearby foreground stellar particles are removed selecting a heliocentric distance greater than 5 kpc. Of the initial 230 stellar particles, only 55 inhabit the same RV$-$\FeH{} region as Sculptor. Of the 55 stellar particles,  only one has a proper motion compatible with Sculptor members. However, its photometry in the G band differs more than a dex from the Sculptor stars at the same (BP $-$ RP) colour. None of the simulated stellar particles can reproduce all the Sculptor properties simultaneously. We confirm Target~1~and~2 as new Sculptor members.

\begin{table*}
\caption[]{Radial velocities, equivalent widths, and metallicities of the targets. The metallicities from the three calibrations are reported, S10 \citep{Starkenburg10}, C13 \citep{Carrera13} and L23 \citep{Longeard23}, and the mean from the three inferences. }
\resizebox{1\textwidth}{!}{
\hspace{-0.6cm}
\begin{tabular}{lcccccccc}
\hline
Target  & RV & EW$_{1}$  & EW$_{2}$ &EW$_{3}$ & \FeH & \FeH & \FeH & $\overline{\FeH}$  \\ 
 &  (km s$^{-1})$ & (m\AA) & (m\AA) & (m\AA)& S10 & C13 & L23 &  \\ \hline 
 Target~1  & $148.6 \pm 2.1 \pm 13.3$ &$940 \pm 41$ & $1900 \pm 105$ & $1640 \pm 31$ & $-1.95 \pm 0.33$ &$ -2.66 \pm 0.37$ &$ -2.08 \pm 0.11$ & $ -2.23 \pm 0.38$  \\ 
Target~2   & $112.4 \pm 2.5 \pm 13.3$ & $912 \pm 46$ & $1999 \pm 86$ & $1462 \pm 35$ &$ -1.93 \pm 0.33$ & $ -2.59 \pm 0.37$ &$ -2.03 \pm 0.09$ & $ -2.18 \pm 0.36$  \\ 
\hline
\end{tabular}}
\label{tab:met}
\end{table*}

\section{Discussion}\label{sec:discussion}
\subsection{New Sculptor members}
The two candidate members have positions, photometry, proper motions, radial velocities, and metallicities compatible with known stars in Sculptor (see Figures~\ref{Fig:onsky}~and~\ref{Fig:rv_met_rh}). These properties cannot be reproduced at the same time by the Besan\c{c}on simulations.  We conclude that the two targets analysed in this work are new members of Sculptor. Therefore, Sculptor extends at least out to a projected elliptical distance of 10 $r_h$, which is a physical distance of $\sim3$ kpc from its centre.

\subsection{Metallicity gradients}
Figure~\ref{Fig:rv_met_rh} shows that the metal-rich population ($\FeH{}\gtrsim -1.5$) is more centrally concentrated and kinematically colder than the metal-poor stars ($\FeH{}\lesssim -1.5$). As a matter of fact,  90 percent of the metal-rich population lies within $2.0 r_h$, while the same fraction of the more metal-poor stars is contained within $3.7 r_h$. This result is in agreement with previous work \citep[\eg][]{Tolstoy04,Battaglia08b,Walker11,Tolstoy23}, in which the metal-rich and centrally concentrated population has a smaller velocity dispersion  ($\sim6-7\kms$) than the more metal-poor and diffuse population ($\sim10-12\kms$). This gradient in metallicity can be explained by an outside-in star formation history or could be the result of an early merger which dispersed the metal poor component before the subsequent formation of the metal-rich stars. \citep[\eg][]{Zhang12,BenitezLlambay16,Revaz18}. While the presence of a gradient in Sculptor can be an effect of mechanisms in play at early times, the presence of stars in the extreme outskirts might be a signature of recent tidal perturbations, as discussed in the following section.

\subsection{Sculptor is affected by tides}
\begin{figure*}
\includegraphics[width=\textwidth]{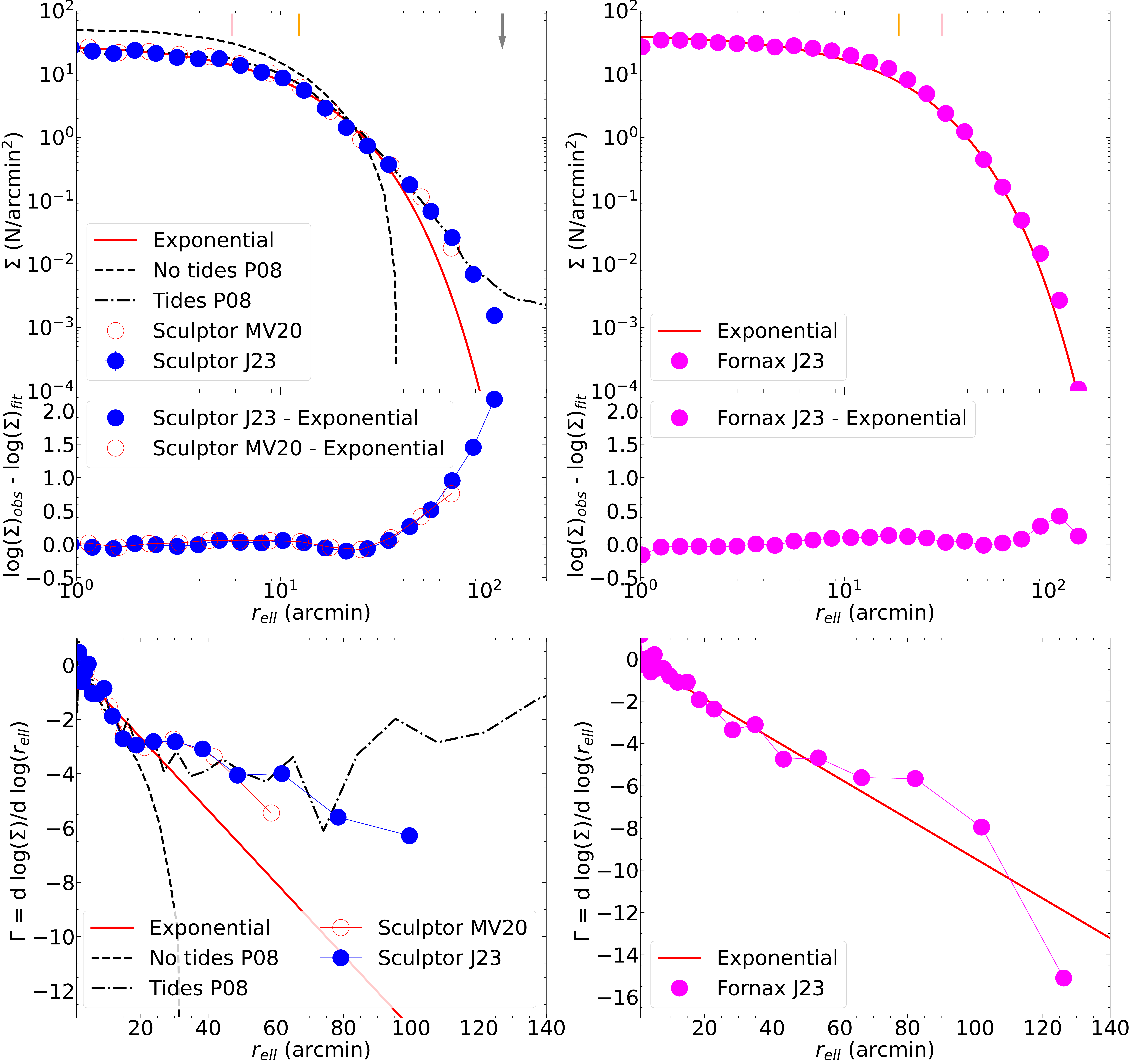}
\caption{Surface number density distribution $\Sigma$ and its spatial logarithmic derivative $\Gamma$. Left panels: Sculptor. Right panels: Fornax. Top and bottom panels: Sculptor candidate members from J23 and MV20 are marked with blue and empty red circles, respectively. Fornax candidate members (magenta circles) are from J23. P08 models are marked with a dashed line (no tide) and with a dash-dotted line (after first apocentric passage model), which are reported only for Sculptor. Red line is the exponential fit to J23 data. Grey arrow marks the position of the two outermost members analysed in this work. Pink and orange ticks denotes the core and the half-light radius of the systems. Central panels: difference between J23 data and the exponential fit. For Sculptor, the exponential fit is able to reproduce the inner surface number density distribution of the MV20 data as well.}
\label{Fig:surf}
\end{figure*}

At face value, the identification of member stars as far out as 10 half-light radii from the centre of Sculptor is rather surprising \citep[see also][]{Qi22}. For an exponential profile, the surface number density of stars at that radius is expected to drop by 6 orders of magnitude from its central value and, therefore, the chance of detecting member stars that far from the centre should be negligible.  Indeed, as shown in the top-left panel of Figure~\ref{Fig:surf}, the stars presented in our study are, at $\sim100$ arcmin, part of a clear excess over an exponential profile (solid red line) fit to the inner regions of Sculptor’s surface number density profile. Because of this excess, the surface number density at 10 half light radii is roughly 100 times higher than expected from an exponential profile.
 
What is the origin of this excess? Is it innate to Sculptor, or is it the result of internal processes that have driven stars out, such as gravitational fluctuations driven by the gas inflows and outflows \citep[\eg][]{ElBadry18b}, or of external processes, such as early accretion or mergers \citep{BenitezLlambay16}? Or is it perhaps driven by the effects of Galactic tides? Although it is difficult to make a definitive case for either scenario without further data or detailed simulation, we believe that the data favours a tidal interpretation, as detailed below.
 
Scl is a satellite of the Milky Way, currently at $\sim86$ kpc from its centre \citep[\eg][]{Mcconnachie12}, and moving out on an orbit with pericentric and apocentric distance of $\sim55$ kpc and $\gtrsim145$ kpc, respectively \citep[][]{Li22,Battaglia22,Pace22}. Despite the relatively large inferred pericentre, Sculptor shows sign of internal velocity gradients, which have been interpreted as a consequence of Galactic tides \citep[see, \eg][]{MartinezGarcia23}. Tides are expected also to leave their imprint on the surface number density profile ($\Sigma$) of a dwarf. This imprint is best identified by considering the logarithmic slope of the profile, $\Gamma=\rm{d}\log\Sigma/\rm{d}\log r$,  which is shown, as a function of radius, in the bottom left panel of Figure~\ref{Fig:surf}. For an exponential profile, $\Gamma$ decreases linearly with radius, as shown by the solid red line, with a slope equal to the inverse of the exponential scale length. 
 
As a result of a tidal perturbation, some stars may gain energy and move to orbits with larger apocentric radii, creating an outer "excess" of stars that resemble that of Sculptor. A signature of this perturbation is a sudden departure of $\Gamma$ from a linear dependence on radius; \ie{} a "kink" in the $\Gamma$ profile like the one clearly seen for Sculptor at r$_{\rm{ell}}\sim25$ arcmin from the centre. Beyond the kink, the tidal excess should approach a power-law of slope $\Gamma=-4$, as expected if stars have been moved to very loosely bound orbits, in effect populating phase space all the way to zero energy, E $\sim 0$ \citep[][]{White87,Jaffe87,Penarrubia09}.
 
This behaviour is illustrated in Figure~\ref{Fig:surf} by the dashed and dot-dashed black curves, which are taken from the tidally-perturbed models of \citet[][see the top-left panel of their Figure~4]{Penarrubia08}. After scaling to the same half-light radius, the tidally perturbed (dot-dashed) profile fits remarkably well the outer excess seen in Sculptor, including the predicted R$^{-4}$ behaviour in the outer regions, even though this simulation was not intended in any way to reproduce Sculptor. 
 
In the simulations, the $R^{-4}$ region ends at a “break” radius where the $\Sigma$ profile flattens and $\Gamma$ increases sharply (this occurs at r$_{\rm{ell}}\sim80$ arcmin in the simulated profile). The location of this "break" corresponds roughly to where the average motion of stars transitions from equilibrium ($\langle v_r \rangle \approx0$) to becoming dominated by outward motion ($\langle v_r \rangle > 0$), and corresponds to the radius where the local crossing time equals the time elapsed since pericenter \citep[see][]{Aguilar86,Navarro90,Penarrubia09}. Outside the break, stars are still moving to their new apocentres and have not yet reached equilibrium.
 
In the case of Sculptor,  assuming that the system last passed  pericentre $\sim0.41$ Gyr ago \citep{Battaglia22}, the break radius is expected\footnote{The relation from \citet{Penarrubia09} is used, r$_b~=~C~\cdot~\sigma_{\rm{v}}~\cdot~t_{\rm{peri}}$, where C is a coefficient ($=0.55$),  $\sigma_{\rm{v}}$ is the velocity dispersion of the system ($9.2\kms$, see Table~\ref{tab:scuprop}), and $t_{\rm{peri}}$ is the time since the last pericentric passage. For the latter, the value derived by \citet{Battaglia22} using the MW potential perturbed by the Large Magellanic Cloud is adopted.} to be at $\sim 2.1$ kpc ($\sim85$ arcmin or $\sim6.9$ r$_h$), very close to the two outermost points of the $\Sigma$ profile\footnote{It is quite difficult to estimate robustly the logarithmic slope of the profile at that radius given current data. A finer binning of the J23 data would show a hint of a plateau in the $\Sigma$ profile around the inferred break radius. However, a finer grid leads to a much higher noise in the $\Gamma$ profile. We defer a more detailed analysis to a future contribution.}. This is a robust prediction of the tidal interpretation that could, in principle, be verified in future work. The "kink" in the $\Gamma$ profile and the $\Gamma=-4$ are not expected in models where the excess stars  in the outskirts are a result of mergers or potential fluctuations \citep[\eg][]{Penarrubia09,BenitezLlambay16}, and clearly favour a tidal interpretation.

Could the outer excess be instead just an innate feature of Sculptor’s density profile rather than a result of a tidal perturbation? If the outer excess is actually tidal in origin, then it should be absent in dwarf spheroidals that have not been affected by tides.  This is the case of Fornax, whose large Galactocentric distance, together with its large orbital pericentre, indicate that tides have played at best a minor role in perturbing its structure \citep[\eg][and references therein]{Borukhovetskaya22,Yang22}. As shown in the right-hand panels of Figure~\ref{Fig:surf}, Fornax shows a profile with little evidence for any excess over the whole radial range plotted here, which spans more than 5 decades in surface number density. 
 
We conclude that the outer excess in the density profile of Sculptor has been driven by tides. In addition, the orbital motion of Target~1~and~2 can be perturbed by the tidal effects. Further work designed to confirm that the tidal signatures described above are consistent with Sculptor's relatively large pericentre, and search for corroborating evidence for this conclusion, such as the presence of a "break radius" in its outer profile, or even the presence of faint tidal tails even further out, is clearly needed.

\section{Conclusions}\label{sec:conclusions}
The two outermost candidate members of Sculptor dwarf galaxy have been selected from the Bayesian algorithm of \citet{Jensen23} and observed with GMOS spectrograph at Gemini South. All the candidate members (J23, MV20) and known Scl members from DART and APOGEE DR17 have been gathered. We find that:
\begin{enumerate}
    \item the two candidates are actually new members given their radial velocity and metallicity.  Therefore, Scl is now known to extend out to a projected elliptical distance of $\sim10$ r$_h$, or $\sim3$ kpc, from its centre (see Figures~\ref{Fig:onsky}~and~\ref{Fig:rv_met_rh});
    \item as noted in earlier work, the radial velocity and metallicity distributions of Sculptor indicate that the most metal-poor stars are more dispersed spatially and kinematically than the more metal-rich population (see Figure~\ref{Fig:rv_met_rh}). This could be explained by an outside-in star formation scenario or an early merger;
    \item the surface number density distribution and its logarithmic derivative (see Figure~\ref{Fig:surf}) suggest that Sculptor has been influenced by tidal effects, starting from a distance of $\sim 25$ arcmin, or $\sim 2$ r$_h$, or $\sim620$ pc from its centre;
    \item tides are likely  responsible for the presence of Target~1~and~2 in the extreme outskirts of Sculptor.
\end{enumerate}

Recent work on the outskirts of ultra faint and dwarf galaxies \citep[][this work]{Chiti21,Filion21,Longeard22,Yang22,Waller23,Sestito23Umi,Longeard23,Jensen23}  has uncovered a new population of stars that would have been undiscoverable without the powerful synergy between the exquisite Gaia data, efficient algorithms to remove foreground stars, and large aperture telescopes.

\section*{Acknowledgements}
We acknowledge and respect the l\textschwa\textvbaraccent {k}$^{\rm w}$\textschwa\ng{}\textschwa n peoples on whose traditional territory the University of Victoria stands and the Songhees, Esquimalt and $\ubar{\rm W}$S\'ANE\'C  peoples whose historical relationships with the land continue to this day.

We thank Nicolas Longeard for sharing the Monte Carlo chain drawings from their metallicity calibration.

We want to thank the supporter astronomers Venu Kalari and Juan Font-Serra for their support during Phase II.

FS thanks the Dr. Margaret "Marmie" Perkins Hess postdoctoral fellowship for funding his work at the University of Victoria. KAV thanks the National Sciences and Engineering Research Council of Canada for funding through the Discovery Grants and CREATE programs. 

This work is based on observations obtained with Gemini South/GMOS, as part of the Gemini Fast Turnaround Program GS-2022B-FT-205. Based on observations obtained at the international Gemini Observatory, a program of NSF’s NOIRLab, which is managed by the Association of Universities for Research in Astronomy (AURA) under a cooperative agreement with the National Science Foundation. On behalf of the Gemini Observatory partnership: the National Science Foundation (United States), National Research Council (Canada), Agencia Nacional de Investigaci\'{o}n y Desarrollo (Chile), Ministerio de Ciencia, Tecnolog\'{i}a e Innovaci\'{o}n (Argentina), Minist\'{e}rio da Ci\^{e}ncia, Tecnologia, Inova\c{c}\~{o}es e Comunica\c{c}\~{o}es (Brazil), and Korea Astronomy and Space Science Institute (Republic of Korea).

This work has made use of data from the European Space Agency (ESA) mission {\it Gaia} (\url{https://www.cosmos.esa.int/gaia}), processed by the {\it Gaia} Data Processing and Analysis Consortium (DPAC, \url{https://www.cosmos.esa.int/web/gaia/dpac/consortium}). Funding for the DPAC has been provided by national institutions, in particular the institutions participating in the {\it Gaia} Multilateral Agreement.

Funding for the Sloan Digital Sky Survey IV has been provided by the Alfred P. Sloan Foundation, the U.S. Department of Energy Office of Science, and the Participating Institutions. SDSS-IV acknowledges support and resources from the Center for High-Performance Computing at the University of Utah. The SDSS web site is \url{www.sdss.org}.

SDSS-IV is managed by the Astrophysical Research Consortium for the Participating Institutions of the SDSS Collaboration including the Brazilian Participation Group, the Carnegie Institution for Science, Carnegie Mellon University, the Chilean Participation Group, the French Participation Group, Harvard-Smithsonian Center for Astrophysics, Instituto de Astrof\'isica de Canarias, The Johns Hopkins University, Kavli Institute for the Physics and Mathematics of the Universe (IPMU) / University of Tokyo, the Korean Participation Group, Lawrence Berkeley National Laboratory, Leibniz Institut f\"ur Astrophysik Potsdam (AIP), Max-Planck-Institut f\"ur Astronomie (MPIA Heidelberg), Max-Planck-Institut f\"ur Astrophysik (MPA Garching), Max-Planck-Institut f\"ur Extraterrestrische Physik (MPE), National Astronomical Observatories of China, New Mexico State University, New York University, University of Notre Dame, Observat\'ario Nacional / MCTI, The Ohio State University, Pennsylvania State University, Shanghai Astronomical Observatory, United Kingdom Participation Group, Universidad Nacional Aut\'onoma de M\'exico, University of Arizona, University of Colorado Boulder, University of Oxford, University of Portsmouth, University of Utah, University of Virginia, University of Washington, University of Wisconsin, Vanderbilt University, and Yale University.

This research has made use of the SIMBAD database, operated at CDS, Strasbourg, France \citep{Wenger00}. This work made extensive use of \textsc{TOPCAT} \citep{Taylor05}.

\section*{Data Availability}
GMOS spectra will be available at the Gemini Archive web page \url{https://archive.gemini.edu/searchform} after the proprietary time. All data are incorporated into the article.

\bibliographystyle{mn2e}
\bibliography{sculptor_gmos}

\begin{thebibliography}{97}
\expandafter\ifx\csname natexlab\endcsname\relax\def\natexlab#1{#1}\fi

\bibitem[{{Abdurro'uf} {et~al}\mbox{.}(2022){Abdurro'uf}, {Accetta}, {Aerts},
  {Silva Aguirre}, {Ahumada}, {Ajgaonkar}, {Filiz Ak}, {Alam}, {Allende
  Prieto}, {Almeida}, \& et~al.}]{APOGEEDR17}
{Abdurro'uf} {et~al.}, 2022, \apjs, 259, 35

\bibitem[{{Aguilar} \& {White}(1986)}]{Aguilar86}
{Aguilar} L.~A., {White} S.~D.~M., 1986, \apj, 307, 97

\bibitem[{{Andrae} {et~al}\mbox{.}(2018){Andrae}, {Fouesneau}, {Creevey},
  {Ordenovic}, {Mary}, {Burlacu}, {Chaoul}, {Jean-Antoine-Piccolo},
  {Kordopatis}, {Korn}, {Lebreton}, {Panem}, {Pichon}, {Th{\'e}venin},
  {Walmsley}, \& {Bailer-Jones}}]{Andrae18}
{Andrae} R. {et~al.}, 2018, \aap, 616, A8

\bibitem[{{Battaglia} {et~al}\mbox{.}(2008){Battaglia}, {Helmi}, {Tolstoy},
  {Irwin}, {Hill}, \& {Jablonka}}]{Battaglia08b}
{Battaglia} G., {Helmi} A., {Tolstoy} E., {Irwin} M., {Hill} V., {Jablonka} P.,
  2008, \apjl, 681, L13

\bibitem[{{Battaglia} {et~al}\mbox{.}(2022){Battaglia}, {Taibi}, {Thomas}, \&
  {Fritz}}]{Battaglia22}
{Battaglia} G., {Taibi} S., {Thomas} G.~F., {Fritz} T.~K., 2022, \aap, 657, A54

\bibitem[{{Ben{\'\i}tez-Llambay} {et~al}\mbox{.}(2016){Ben{\'\i}tez-Llambay},
  {Navarro}, {Abadi}, {Gottl{\"o}ber}, {Yepes}, {Hoffman}, \&
  {Steinmetz}}]{BenitezLlambay16}
{Ben{\'\i}tez-Llambay} A., {Navarro} J.~F., {Abadi} M.~G., {Gottl{\"o}ber} S.,
  {Yepes} G., {Hoffman} Y., {Steinmetz} M., 2016, \mnras, 456, 1185

\bibitem[{{Borukhovetskaya} {et~al}\mbox{.}(2022){Borukhovetskaya}, {Errani},
  {Navarro}, {Fattahi}, \& {Santos-Santos}}]{Borukhovetskaya22}
{Borukhovetskaya} A., {Errani} R., {Navarro} J.~F., {Fattahi} A.,
  {Santos-Santos} I., 2022, \mnras, 509, 5330

\bibitem[{{Bressan} {et~al}\mbox{.}(2012){Bressan}, {Marigo}, {Girardi},
  {Salasnich}, {Dal Cero}, {Rubele}, \& {Nanni}}]{Bressan12}
{Bressan} A., {Marigo} P., {Girardi} L., {Salasnich} B., {Dal Cero} C.,
  {Rubele} S., {Nanni} A., 2012, \mnras, 427, 127

\bibitem[{{Bullock} \& {Boylan-Kolchin}(2017)}]{Bullock17}
{Bullock} J.~S., {Boylan-Kolchin} M., 2017, \araa, 55, 343

\bibitem[{{Carrera} {et~al}\mbox{.}(2013){Carrera}, {Pancino}, {Gallart}, \&
  {del Pino}}]{Carrera13}
{Carrera} R., {Pancino} E., {Gallart} C., {del Pino} A., 2013, \mnras, 434,
  1681

\bibitem[{Chambers {et~al}\mbox{.}(2016)Chambers, Magnier, Metcalfe,
  Flewelling, Huber, Waters, Denneau, Draper, Farrow, Finkbeiner, Holmberg,
  Koppenhoefer, Price, Rest, Saglia, Schlafly, Smartt, Sweeney, Wainscoat,
  Burgett, Chastel, Grav, Heasley, Hodapp, Jedicke, Kaiser, Kudritzki, Luppino,
  Lupton, Monet, Morgan, Onaka, Shiao, Stubbs, Tonry, White, Ba{\~n}ados, Bell,
  Bender, Bernard, Boegner, Boffi, Botticella, Calamida, Casertano, Chen, Chen,
  Cole, Deacon, Frenk, Fitzsimmons, Gezari, Gibbs, Goessl, Goggia, Gourgue,
  Goldman, Grant, Grebel, Hambly, Hasinger, Heavens, Heckman, Henderson,
  Henning, Holman, Hopp, Ip, Isani, Jackson, Keyes, Koekemoer, Kotak, Le,
  Liska, Long, Lucey, Liu, Martin, Masci, McLean, Mindel, Misra, Morganson,
  Murphy, Obaika, Narayan, Nieto-Santisteban, Norberg, Peacock, Pier, Postman,
  Primak, Rae, Rai, Riess, Riffeser, Rix, R{\"o}ser, Russel, Rutz, Schilbach,
  Schultz, Scolnic, Strolger, Szalay, Seitz, Small, Smith, Soderblom, Taylor,
  Thomson, Taylor, Thakar, Thiel, Thilker, Unger, Urata, Valenti, Wagner,
  Walder, Walter, Watters, Werner, Wood-Vasey, \& Wyse}]{Chambers16}
Chambers K.~C. {et~al.}, 2016, The pan-starrs1 surveys

\bibitem[{{Chiti} {et~al}\mbox{.}(2021){Chiti}, {Frebel}, {Simon}, {Erkal},
  {Chang}, {Necib}, {Ji}, {Jerjen}, {Kim}, \& {Norris}}]{Chiti21}
{Chiti} A. {et~al.}, 2021, Nature Astronomy, 5, 392

\bibitem[{{Deason}, {Wetzel} \& {Garrison-Kimmel}(2014){Deason}, {Wetzel}, \&
  {Garrison-Kimmel}}]{Deason14}
{Deason} A., {Wetzel} A., {Garrison-Kimmel} S., 2014, \apj, 794, 115

\bibitem[{{Deason} {et~al}\mbox{.}(2022){Deason}, {Bose}, {Fattahi},
  {Amorisco}, {Hellwing}, \& {Frenk}}]{Deason22}
{Deason} A.~J., {Bose} S., {Fattahi} A., {Amorisco} N.~C., {Hellwing} W.,
  {Frenk} C.~S., 2022, \mnras, 511, 4044

\bibitem[{{Dey} {et~al}\mbox{.}(2019){Dey}, {Schlegel}, {Lang}, {Blum},
  {Burleigh}, {Fan}, {Findlay}, {Finkbeiner}, {Herrera}, {Juneau}, {Landriau},
  {Levi}, {McGreer}, {Meisner}, {Myers}, {Moustakas}, {Nugent}, {Patej},
  {Schlafly}, {Walker}, {Valdes}, {Weaver}, {Y{\`e}che}, {Zou}, {Zhou},
  {Abareshi}, {Abbott}, {Abolfathi}, {Aguilera}, {Alam}, {Allen}, {Alvarez},
  {Annis}, {Ansarinejad}, {Aubert}, {Beechert}, {Bell}, {BenZvi}, {Beutler},
  {Bielby}, {Bolton}, {Brice{\~n}o}, {Buckley-Geer}, {Butler}, {Calamida},
  {Carlberg}, {Carter}, {Casas}, {Castander}, {Choi}, {Comparat},
  {Cukanovaite}, {Delubac}, {DeVries}, {Dey}, {Dhungana}, {Dickinson}, {Ding},
  {Donaldson}, {Duan}, {Duckworth}, {Eftekharzadeh}, {Eisenstein}, {Etourneau},
  {Fagrelius}, {Farihi}, {Fitzpatrick}, {Font-Ribera}, {Fulmer},
  {G{\"a}nsicke}, {Gaztanaga}, {George}, {Gerdes}, {Gontcho}, {Gorgoni},
  {Green}, {Guy}, {Harmer}, {Hernandez}, {Honscheid}, {Huang}, {James},
  {Jannuzi}, {Jiang}, {Joyce}, {Karcher}, {Karkar}, {Kehoe}, {Kneib},
  {Kueter-Young}, {Lan}, {Lauer}, {Le Guillou}, {Le Van Suu}, {Lee}, {Lesser},
  {Perreault Levasseur}, {Li}, {Mann}, {Marshall}, {Mart{\'\i}nez-V{\'a}zquez},
  {Martini}, {du Mas des Bourboux}, {McManus}, {Meier}, {M{\'e}nard},
  {Metcalfe}, {Mu{\~n}oz-Guti{\'e}rrez}, {Najita}, {Napier}, {Narayan},
  {Newman}, {Nie}, {Nord}, {Norman}, {Olsen}, {Paat}, {Palanque-Delabrouille},
  {Peng}, {Poppett}, {Poremba}, {Prakash}, {Rabinowitz}, {Raichoor}, {Rezaie},
  {Robertson}, {Roe}, {Ross}, {Ross}, {Rudnick}, {Safonova}, {Saha},
  {S{\'a}nchez}, {Savary}, {Schweiker}, {Scott}, {Seo}, {Shan}, {Silva},
  {Slepian}, {Soto}, {Sprayberry}, {Staten}, {Stillman}, {Stupak}, {Summers},
  {Sien Tie}, {Tirado}, {Vargas-Maga{\~n}a}, {Vivas}, {Wechsler}, {Williams},
  {Yang}, {Yang}, {Yapici}, {Zaritsky}, {Zenteno}, {Zhang}, {Zhang}, {Zhou}, \&
  {Zhou}}]{DESI19}
{Dey} A. {et~al.}, 2019, \aj, 157, 168

\bibitem[{{Drlica-Wagner} {et~al}\mbox{.}(2021){Drlica-Wagner}, {Carlin},
  {Nidever}, {Ferguson}, {Kuropatkin}, {Adam{\'o}w}, {Cerny}, {Choi},
  {Esteves}, {Mart{\'\i}nez-V{\'a}zquez}, {Mau}, {Miller}, {Mutlu-Pakdil},
  {Neilsen}, {Olsen}, {Pace}, {Riley}, {Sakowska}, {Sand}, {Santana-Silva},
  {Tollerud}, {Tucker}, {Vivas}, {Zaborowski}, {Zenteno}, {Abbott}, {Allam},
  {Bechtol}, {Bell}, {Bell}, {Bilaji}, {Bom}, {Carballo-Bello},
  {Crnojevi{\'c}}, {Cioni}, {Diaz-Ocampo}, {de Boer}, {Erkal}, {Gruendl},
  {Hernandez-Lang}, {Hughes}, {James}, {Johnson}, {Li}, {Mao},
  {Mart{\'\i}nez-Delgado}, {Massana}, {McNanna}, {Morgan}, {Nadler},
  {No{\"e}l}, {Palmese}, {Peter}, {Rykoff}, {S{\'a}nchez}, {Shipp}, {Simon},
  {Smercina}, {Soares-Santos}, {Stringfellow}, {Tavangar}, {van der Marel},
  {Walker}, {Wechsler}, {Wu}, {Yanny}, {Fitzpatrick}, {Huang}, {Jacques},
  {Nikutta}, {Scott}, \& {Astro Data Lab}}]{DELVE21}
{Drlica-Wagner} A. {et~al.}, 2021, \apjs, 256, 2

\bibitem[{{El-Badry} {et~al}\mbox{.}(2018){El-Badry}, {Bradford}, {Quataert},
  {Geha}, {Boylan-Kolchin}, {Weisz}, {Wetzel}, {Hopkins}, {Chan}, {Fitts},
  {Kere{\v{s}}}, \& {Faucher-Gigu{\`e}re}}]{ElBadry18b}
{El-Badry} K. {et~al.}, 2018, \mnras, 477, 1536

\bibitem[{{Evans} {et~al}\mbox{.}(2018){Evans}, {Riello}, {De Angeli},
  {Carrasco}, {Montegriffo}, {Fabricius}, {Jordi}, {Palaversa}, {Diener},
  {Busso}, {Cacciari}, {van Leeuwen}, {Burgess}, {Davidson}, {Harrison},
  {Hodgkin}, {Pancino}, {Richards}, {Altavilla}, {Balaguer-N{\'u}{\~n}ez},
  {Barstow}, {Bellazzini}, {Brown}, {Castellani}, {Cocozza}, {De Luise},
  {Delgado}, {Ducourant}, {Galleti}, {Gilmore}, {Giuffrida}, {Holl}, {Kewley},
  {Koposov}, {Marinoni}, {Marrese}, {Osborne}, {Piersimoni}, {Portell},
  {Pulone}, {Ragaini}, {Sanna}, {Terrett}, {Walton}, {Wevers}, \&
  {Wyrzykowski}}]{Evans18}
{Evans} D.~W. {et~al.}, 2018, \aap, 616, A4

\bibitem[{{Fattahi} {et~al}\mbox{.}(2018){Fattahi}, {Navarro}, {Frenk}, {Oman},
  {Sawala}, \& {Schaller}}]{Fattahi18}
{Fattahi} A., {Navarro} J.~F., {Frenk} C.~S., {Oman} K.~A., {Sawala} T.,
  {Schaller} M., 2018, \mnras, 476, 3816

\bibitem[{{Fenner} {et~al}\mbox{.}(2006){Fenner}, {Gibson}, {Gallino}, \&
  {Lugaro}}]{Fenner06}
{Fenner} Y., {Gibson} B.~K., {Gallino} R., {Lugaro} M., 2006, \apj, 646, 184

\bibitem[{{Filion} \& {Wyse}(2021)}]{Filion21}
{Filion} C., {Wyse} R. F.~G., 2021, \apj, 923, 218

\bibitem[{{Frebel}, {Kirby} \& {Simon}(2010){Frebel}, {Kirby}, \&
  {Simon}}]{Frebel10}
{Frebel} A., {Kirby} E.~N., {Simon} J.~D., 2010, \nat, 464, 72

\bibitem[{{Frenk} {et~al}\mbox{.}(1988){Frenk}, {White}, {Davis}, \&
  {Efstathiou}}]{Frenk88}
{Frenk} C.~S., {White} S. D.~M., {Davis} M., {Efstathiou} G., 1988, \apj, 327,
  507

\bibitem[{{Gaia Collaboration} {et~al}\mbox{.}(2021){Gaia Collaboration},
  {Brown}, {Vallenari}, {Prusti}, {de Bruijne}, {Babusiaux}, {Biermann},
  {Creevey}, {Evans}, {Eyer}, {Hutton}, {Jansen}, {Jordi}, {Klioner},
  {Lammers}, {Lindegren}, {Luri}, {Mignard}, {Panem}, {Pourbaix}, {Randich},
  {Sartoretti}, {Soubiran}, {Walton}, {Arenou}, {Bailer-Jones}, {Bastian},
  {Cropper}, {Drimmel}, {Katz}, {Lattanzi}, {van Leeuwen}, {Bakker},
  {Cacciari}, {Casta{\~n}eda}, {De Angeli}, {Ducourant}, {Fabricius},
  {Fouesneau}, {Fr{\'e}mat}, {Guerra}, {Guerrier}, {Guiraud}, {Jean-Antoine
  Piccolo}, {Masana}, {Messineo}, {Mowlavi}, {Nicolas}, {Nienartowicz},
  {Pailler}, {Panuzzo}, {Riclet}, {Roux}, {Seabroke}, {Sordo}, {Tanga},
  {Th{\'e}venin}, {Gracia-Abril}, {Portell}, {Teyssier}, {Altmann}, {Andrae},
  {Bellas-Velidis}, {Benson}, {Berthier}, {Blomme}, {Brugaletta}, {Burgess},
  {Busso}, {Carry}, {Cellino}, {Cheek}, {Clementini}, {Damerdji}, {Davidson},
  {Delchambre}, {Dell'Oro}, {Fern{\'a}ndez-Hern{\'a}ndez}, {Galluccio},
  {Garc{\'\i}a-Lario}, {Garcia-Reinaldos}, {Gonz{\'a}lez-N{\'u}{\~n}ez},
  {Gosset}, {Haigron}, {Halbwachs}, {Hambly}, {Harrison}, {Hatzidimitriou},
  {Heiter}, {Hern{\'a}ndez}, {Hestroffer}, {Hodgkin}, {Holl}, {Jan{\ss}en},
  {Jevardat de Fombelle}, {Jordan}, {Krone-Martins}, {Lanzafame},
  {L{\"o}ffler}, {Lorca}, {Manteiga}, {Marchal}, {Marrese}, {Moitinho}, {Mora},
  {Muinonen}, {Osborne}, {Pancino}, {Pauwels}, {Petit}, {Recio-Blanco},
  {Richards}, {Riello}, {Rimoldini}, {Robin}, {Roegiers}, {Rybizki}, {Sarro},
  {Siopis}, {Smith}, {Sozzetti}, {Ulla}, {Utrilla}, {van Leeuwen}, {van
  Reeven}, {Abbas}, {Abreu Aramburu}, {Accart}, {Aerts}, {Aguado}, {Ajaj},
  {Altavilla}, {{\'A}lvarez}, {{\'A}lvarez Cid-Fuentes}, {Alves}, {Anderson},
  {Anglada Varela}, {Antoja}, {Audard}, {Baines}, {Baker},
  {Balaguer-N{\'u}{\~n}ez}, {Balbinot}, {Balog}, {Barache}, {Barbato},
  {Barros}, {Barstow}, {Bartolom{\'e}}, {Bassilana}, {Bauchet},
  {Baudesson-Stella}, {Becciani}, {Bellazzini}, {Bernet}, {Bertone}, {Bianchi},
  {Blanco-Cuaresma}, {Boch}, {Bombrun}, {Bossini}, {Bouquillon}, {Bragaglia},
  {Bramante}, {Breedt}, {Bressan}, {Brouillet}, {Bucciarelli}, {Burlacu},
  {Busonero}, {Butkevich}, {Buzzi}, {Caffau}, {Cancelliere}, {C{\'a}novas},
  {Cantat-Gaudin}, {Carballo}, {Carlucci}, {Carnerero}, {Carrasco},
  {Casamiquela}, {Castellani}, {Castro-Ginard}, {Castro Sampol}, {Chaoul},
  {Charlot}, {Chemin}, {Chiavassa}, {Cioni}, {Comoretto}, {Cooper}, {Cornez},
  {Cowell}, {Crifo}, {Crosta}, {Crowley}, {Dafonte}, {Dapergolas}, {David},
  {David}, {de Laverny}, {De Luise}, {De March}, {De Ridder}, {de Souza}, {de
  Teodoro}, {de Torres}, {del Peloso}, {del Pozo}, {Delbo}, {Delgado},
  {Delgado}, {Delisle}, {Di Matteo}, {Diakite}, {Diener}, {Distefano},
  {Dolding}, {Eappachen}, {Edvardsson}, {Enke}, {Esquej}, {Fabre}, {Fabrizio},
  {Faigler}, {Fedorets}, {Fernique}, {Fienga}, {Figueras}, {Fouron},
  {Fragkoudi}, {Fraile}, {Franke}, {Gai}, {Garabato}, {Garcia-Gutierrez},
  {Garc{\'\i}a-Torres}, {Garofalo}, {Gavras}, {Gerlach}, {Geyer}, {Giacobbe},
  {Gilmore}, {Girona}, {Giuffrida}, {Gomel}, {Gomez}, {Gonzalez-Santamaria},
  {Gonz{\'a}lez-Vidal}, {Granvik}, {Guti{\'e}rrez-S{\'a}nchez}, {Guy},
  {Hauser}, {Haywood}, {Helmi}, {Hidalgo}, {Hilger}, {H{\l}adczuk}, {Hobbs},
  {Holland}, {Huckle}, {Jasniewicz}, {Jonker}, {Juaristi Campillo}, {Julbe},
  {Karbevska}, {Kervella}, {Khanna}, {Kochoska}, {Kontizas}, {Kordopatis},
  {Korn}, {Kostrzewa-Rutkowska}, {Kruszy{\'n}ska}, {Lambert}, {Lanza}, {Lasne},
  {Le Campion}, {Le Fustec}, {Lebreton}, {Lebzelter}, {Leccia}, {Leclerc},
  {Lecoeur-Taibi}, {Liao}, {Licata}, {Lindstr{\o}m}, {Lister}, {Livanou},
  {Lobel}, {Madrero Pardo}, {Managau}, {Mann}, {Marchant}, {Marconi}, {Marcos
  Santos}, {Marinoni}, {Marocco}, {Marshall}, {Martin Polo},
  {Mart{\'\i}n-Fleitas}, {Masip}, {Massari}, {Mastrobuono-Battisti}, {Mazeh},
  {McMillan}, {Messina}, {Michalik}, {Millar}, {Mints}, {Molina}, {Molinaro},
  {Moln{\'a}r}, {Montegriffo}, {Mor}, {Morbidelli}, {Morel}, {Morris},
  {Mulone}, {Munoz}, {Muraveva}, {Murphy}, {Musella}, {Noval}, {Ord{\'e}novic},
  {Orr{\`u}}, {Osinde}, {Pagani}, {Pagano}, {Palaversa}, {Palicio}, {Panahi},
  {Pawlak}, {Pe{\~n}alosa Esteller}, {Penttil{\"a}}, {Piersimoni}, {Pineau},
  {Plachy}, {Plum}, {Poggio}, {Poretti}, {Poujoulet}, {Pr{\v{s}}a}, {Pulone},
  {Racero}, {Ragaini}, {Rainer}, {Raiteri}, {Rambaux}, {Ramos}, {Ramos-Lerate},
  {Re Fiorentin}, {Regibo}, {Reyl{\'e}}, {Ripepi}, {Riva}, {Rixon}, {Robichon},
  {Robin}, {Roelens}, {Rohrbasser}, {Romero-G{\'o}mez}, {Rowell}, {Royer},
  {Rybicki}, {Sadowski}, {Sagrist{\`a} Sell{\'e}s}, {Sahlmann}, {Salgado},
  {Salguero}, {Samaras}, {Sanchez Gimenez}, {Sanna}, {Santove{\~n}a},
  {Sarasso}, {Schultheis}, {Sciacca}, {Segol}, {Segovia}, {S{\'e}gransan},
  {Semeux}, {Shahaf}, {Siddiqui}, {Siebert}, {Siltala}, {Slezak}, {Smart},
  {Solano}, {Solitro}, {Souami}, {Souchay}, {Spagna}, {Spoto}, {Steele},
  {Steidelm{\"u}ller}, {Stephenson}, {S{\"u}veges}, {Szabados}, {Szegedi-Elek},
  {Taris}, {Tauran}, {Taylor}, {Teixeira}, {Thuillot}, {Tonello}, {Torra},
  {Torra}, {Turon}, {Unger}, {Vaillant}, {van Dillen}, {Vanel}, {Vecchiato},
  {Viala}, {Vicente}, {Voutsinas}, {Weiler}, {Wevers}, {Wyrzykowski}, {Yoldas},
  {Yvard}, {Zhao}, {Zorec}, {Zucker}, {Zurbach}, \& {Zwitter}}]{GaiaEDR3}
{Gaia Collaboration} {et~al.}, 2021, \aap, 649, A1

\bibitem[{{Genel} {et~al}\mbox{.}(2010){Genel}, {Bouch{\'e}}, {Naab},
  {Sternberg}, \& {Genzel}}]{Genel10}
{Genel} S., {Bouch{\'e}} N., {Naab} T., {Sternberg} A., {Genzel} R., 2010,
  \apj, 719, 229

\bibitem[{{Gimeno} {et~al}\mbox{.}(2016){Gimeno}, {Roth}, {Chiboucas}, {Hibon},
  {Boucher}, {White}, {Rippa}, {Labrie}, {Turner}, {Hanna}, {Lazo},
  {P{\'e}rez}, {Rogers}, {Rojas}, {Placco}, \& {Murowinski}}]{Gimeno16}
{Gimeno} G. {et~al.}, 2016, in Society of Photo-Optical Instrumentation
  Engineers (SPIE) Conference Series, Vol. 9908, Ground-based and Airborne
  Instrumentation for Astronomy VI, {Evans} C.~J., {Simard} L., {Takami} H.,
  eds., p. 99082S

\bibitem[{{Gonz{\'a}lez Hern{\'a}ndez} \& {Bonifacio}(2009)}]{Gonzalez09}
{Gonz{\'a}lez Hern{\'a}ndez} J.~I., {Bonifacio} P., 2009, \aap, 497, 497

\bibitem[{{Grebel}, {Gallagher} \& {Harbeck}(2003){Grebel}, {Gallagher}, \&
  {Harbeck}}]{Grebel03}
{Grebel} E.~K., {Gallagher}, John~S. I., {Harbeck} D., 2003, \aj, 125, 1926

\bibitem[{{Gustafsson} {et~al}\mbox{.}(2008){Gustafsson}, {Edvardsson},
  {Eriksson}, {J{\o}rgensen}, {Nordlund}, \& {Plez}}]{Gustafsson08}
{Gustafsson} B., {Edvardsson} B., {Eriksson} K., {J{\o}rgensen} U.~G.,
  {Nordlund} {\r{A}}., {Plez} B., 2008, \aap, 486, 951

\bibitem[{{Hartwig} {et~al}\mbox{.}(2023){Hartwig}, {Ishigaki}, {Kobayashi},
  {Tominaga}, \& {Nomoto}}]{Hartwig23}
{Hartwig} T., {Ishigaki} M.~N., {Kobayashi} C., {Tominaga} N., {Nomoto} K.,
  2023, \apj, 946, 20

\bibitem[{{Higgs} {et~al}\mbox{.}(2021){Higgs}, {McConnachie}, {Annau},
  {Irwin}, {Battaglia}, {C{\^o}t{\'e}}, {Lewis}, \& {Venn}}]{Higgs21}
{Higgs} C.~R., {McConnachie} A.~W., {Annau} N., {Irwin} M., {Battaglia} G.,
  {C{\^o}t{\'e}} P., {Lewis} G.~F., {Venn} K., 2021, \mnras, 503, 176

\bibitem[{{Hill} {et~al}\mbox{.}(2019){Hill}, {Sk{\'u}lad{\'o}ttir}, {Tolstoy},
  {Venn}, {Shetrone}, {Jablonka}, {Primas}, {Battaglia}, {de Boer},
  {Fran{\c{c}}ois}, {Helmi}, {Kaufer}, {Letarte}, {Starkenburg}, \&
  {Spite}}]{Hill19}
{Hill} V. {et~al.}, 2019, \aap, 626, A15

\bibitem[{{Hook} {et~al}\mbox{.}(2004){Hook}, {J{\o}rgensen},
  {Allington-Smith}, {Davies}, {Metcalfe}, {Murowinski}, \&
  {Crampton}}]{Hook04}
{Hook} I.~M., {J{\o}rgensen} I., {Allington-Smith} J.~R., {Davies} R.~L.,
  {Metcalfe} N., {Murowinski} R.~G., {Crampton} D., 2004, \pasp, 116, 425

\bibitem[{{Ishigaki} {et~al}\mbox{.}(2018){Ishigaki}, {Tominaga}, {Kobayashi},
  \& {Nomoto}}]{Ishigaki18}
{Ishigaki} M.~N., {Tominaga} N., {Kobayashi} C., {Nomoto} K., 2018, \apj, 857,
  46

\bibitem[{{Jablonka} {et~al}\mbox{.}(2015){Jablonka}, {North}, {Mashonkina},
  {Hill}, {Revaz}, {Shetrone}, {Starkenburg}, {Irwin}, {Tolstoy}, {Battaglia},
  {Venn}, {Helmi}, {Primas}, \& {Fran{\c{c}}ois}}]{Jablonka15}
{Jablonka} P. {et~al.}, 2015, \aap, 583, A67

\bibitem[{{Jaffe}(1987)}]{Jaffe87}
{Jaffe} W., 1987, in Structure and Dynamics of Elliptical Galaxies, {de Zeeuw}
  P.~T., ed., Vol. 127, p. 511

\bibitem[{{Jensen et al.}(2023, in prep.)}]{Jensen23}
{Jensen et al.}, 2023, in prep.

\bibitem[{{Kazantzidis} {et~al}\mbox{.}(2011){Kazantzidis}, {{\L}okas},
  {Callegari}, {Mayer}, \& {Moustakas}}]{Kazantzidis11}
{Kazantzidis} S., {{\L}okas} E.~L., {Callegari} S., {Mayer} L., {Moustakas}
  L.~A., 2011, \apj, 726, 98

\bibitem[{{Kielty} {et~al}\mbox{.}(2021){Kielty}, {Venn}, {Sestito},
  {Starkenburg}, {Martin}, {Aguado}, {Arentsen}, {Fabbro}, {Gonz{\'a}lez
  Hern{\'a}ndez}, {Hill}, {Jablonka}, {Lardo}, {Mashonkina}, {Navarro},
  {Sneden}, {Thomas}, {Youakim}, {Bialek}, \& {S{\'a}nchez-Janssen}}]{Kielty21}
{Kielty} C.~L. {et~al.}, 2021, \mnras, 506, 1438

\bibitem[{{Lardo} {et~al}\mbox{.}(2021){Lardo}, {Mashonkina}, {Jablonka},
  {Bonifacio}, {Caffau}, {Aguado}, {Gonz{\'a}lez Hern{\'a}ndez}, {Sestito},
  {Kielty}, {Venn}, {Hill}, {Starkenburg}, {Martin}, {Sitnova}, {Arentsen},
  {Carlberg}, {Navarro}, \& {Kordopatis}}]{Lardo21}
{Lardo} C. {et~al.}, 2021, \mnras, 508, 3068

\bibitem[{{Li} {et~al}\mbox{.}(2022){Li}, {Ji}, {Pace}, {Erkal}, {Koposov},
  {Shipp}, {Da Costa}, {Cullinane}, {Kuehn}, {Lewis}, {Mackey}, {Simpson},
  {Zucker}, {Ferguson}, {Martell}, {Bland-Hawthorn}, {Balbinot}, {Tavangar},
  {Drlica-Wagner}, {De Silva}, \& {Simon}}]{Li22}
{Li} T.~S. {et~al.}, 2022, \apj, 928, 30

\bibitem[{{Longeard} {et~al}\mbox{.}(2022){Longeard}, {Jablonka}, {Arentsen},
  {Thomas}, {Aguado}, {Carlberg}, {Lucchesi}, {Malhan}, {Martin},
  {McConnachie}, {Navarro}, {S{\'a}nchez-Janssen}, {Sestito}, {Starkenburg}, \&
  {Yuan}}]{Longeard22}
{Longeard} N. {et~al.}, 2022, \mnras, 516, 2348

\bibitem[{{Longeard} {et~al}\mbox{.}(2023){Longeard}, {Jablonka}, {Battaglia},
  {Malhan}, {Martin}, {S{\'a}nchez-Janssen}, {Sestito}, {Starkenburg}, \&
  {Venn}}]{Longeard23}
---, 2023, arXiv e-prints, arXiv:2304.13046

\bibitem[{{Marigo} {et~al}\mbox{.}(2008){Marigo}, {Girardi}, {Bressan},
  {Groenewegen}, {Silva}, \& {Granato}}]{Marigo08}
{Marigo} P., {Girardi} L., {Bressan} A., {Groenewegen} M.~A.~T., {Silva} L.,
  {Granato} G.~L., 2008, \aap, 482, 883

\bibitem[{{Martin} {et~al}\mbox{.}(2009){Martin}, {McConnachie}, {Irwin},
  {Widrow}, {Ferguson}, {Ibata}, {Dubinski}, {Babul}, {Chapman}, {Fardal},
  {Lewis}, {Navarro}, \& {Rich}}]{Martin09}
{Martin} N.~F. {et~al.}, 2009, \apj, 705, 758

\bibitem[{{Mart{\'\i}nez-Garc{\'\i}a}, {del Pino} \&
  {Aparicio}(2023){Mart{\'\i}nez-Garc{\'\i}a}, {del Pino}, \&
  {Aparicio}}]{MartinezGarcia23}
{Mart{\'\i}nez-Garc{\'\i}a} A.~M., {del Pino} A., {Aparicio} A., 2023, \mnras,
  518, 3083

\bibitem[{{McConnachie}(2012)}]{Mcconnachie12}
{McConnachie} A.~W., 2012, \aj, 144, 4

\bibitem[{{McConnachie} {et~al}\mbox{.}(2009){McConnachie}, {Irwin}, {Ibata},
  {Dubinski}, {Widrow}, {Martin}, {C{\^o}t{\'e}}, {Dotter}, {Navarro},
  {Ferguson}, {Puzia}, {Lewis}, {Babul}, {Barmby}, {Bienaym{\'e}}, {Chapman},
  {Cockcroft}, {Collins}, {Fardal}, {Harris}, {Huxor}, {Mackey},
  {Pe{\~n}arrubia}, {Rich}, {Richer}, {Siebert}, {Tanvir}, {Valls-Gabaud}, \&
  {Venn}}]{McConnachie09}
{McConnachie} A.~W. {et~al.}, 2009, \nat, 461, 66

\bibitem[{{McConnachie} \& {Venn}(2020{\natexlab{a}})}]{McVenn2020a}
{McConnachie} A.~W., {Venn} K.~A., 2020{\natexlab{a}}, \aj, 160, 124

\bibitem[{{McConnachie} \& {Venn}(2020{\natexlab{b}})}]{McVenn2020b}
---, 2020{\natexlab{b}}, Research Notes of the American Astronomical Society,
  4, 229

\bibitem[{{Moster}, {Naab} \& {White}(2013){Moster}, {Naab}, \&
  {White}}]{Moster13}
{Moster} B.~P., {Naab} T., {White} S. D.~M., 2013, \mnras, 428, 3121

\bibitem[{{Mucciarelli}, {Bellazzini} \& {Massari}(2021){Mucciarelli},
  {Bellazzini}, \& {Massari}}]{Mucciarelli21}
{Mucciarelli} A., {Bellazzini} M., {Massari} D., 2021, \aap, 653, A90

\bibitem[{{Navarro}(1990)}]{Navarro90}
{Navarro} J.~F., 1990, \mnras, 242, 311

\bibitem[{{Navarro}, {Frenk} \& {White}(1997){Navarro}, {Frenk}, \&
  {White}}]{NavarroFrenkWhite97}
{Navarro} J.~F., {Frenk} C.~S., {White} S.~D.~M., 1997, \apj, 490, 493

\bibitem[{{Pace}, {Erkal} \& {Li}(2022){Pace}, {Erkal}, \& {Li}}]{Pace22}
{Pace} A.~B., {Erkal} D., {Li} T.~S., 2022, \apj, 940, 136

\bibitem[{{Pe{\~n}arrubia}, {Navarro} \& {McConnachie}(2008){Pe{\~n}arrubia},
  {Navarro}, \& {McConnachie}}]{Penarrubia08}
{Pe{\~n}arrubia} J., {Navarro} J.~F., {McConnachie} A.~W., 2008, \apj, 673, 226

\bibitem[{{Pe{\~n}arrubia} {et~al}\mbox{.}(2009){Pe{\~n}arrubia}, {Navarro},
  {McConnachie}, \& {Martin}}]{Penarrubia09}
{Pe{\~n}arrubia} J., {Navarro} J.~F., {McConnachie} A.~W., {Martin} N.~F.,
  2009, \apj, 698, 222

\bibitem[{{Placco} {et~al}\mbox{.}(2021){Placco}, {Sneden}, {Roederer},
  {Lawler}, {Den Hartog}, {Hejazi}, {Maas}, \& {Bernath}}]{Placco21}
{Placco} V.~M., {Sneden} C., {Roederer} I.~U., {Lawler} J.~E., {Den Hartog}
  E.~A., {Hejazi} N., {Maas} Z., {Bernath} P., 2021, Research Notes of the
  American Astronomical Society, 5, 92

\bibitem[{{Plez}(2012)}]{Plez12}
{Plez} B., 2012, {Turbospectrum: Code for spectral synthesis}

\bibitem[{{Qi} {et~al}\mbox{.}(2022){Qi}, {Zivick}, {Pace}, {Riley}, \&
  {Strigari}}]{Qi22}
{Qi} Y., {Zivick} P., {Pace} A.~B., {Riley} A.~H., {Strigari} L.~E., 2022,
  \mnras, 512, 5601

\bibitem[{{Revaz} \& {Jablonka}(2018)}]{Revaz18}
{Revaz} Y., {Jablonka} P., 2018, \aap, 616, A96

\bibitem[{{Riello} {et~al}\mbox{.}(2021){Riello}, {De Angeli}, {Evans},
  {Montegriffo}, {Carrasco}, {Busso}, {Palaversa}, {Burgess}, {Diener},
  {Davidson}, {Rowell}, {Fabricius}, {Jordi}, {Bellazzini}, {Pancino},
  {Harrison}, {Cacciari}, {van Leeuwen}, {Hambly}, {Hodgkin}, {Osborne},
  {Altavilla}, {Barstow}, {Brown}, {Castellani}, {Cowell}, {De Luise},
  {Gilmore}, {Giuffrida}, {Hidalgo}, {Holland}, {Marinoni}, {Pagani},
  {Piersimoni}, {Pulone}, {Ragaini}, {Rainer}, {Richards}, {Sanna}, {Walton},
  {Weiler}, \& {Yoldas}}]{Riello21}
{Riello} M. {et~al.}, 2021, \aap, 649, A3

\bibitem[{{Robin} {et~al}\mbox{.}(2017){Robin}, {Bienaym{\'e}},
  {Fern{\'a}ndez-Trincado}, \& {Reyl{\'e}}}]{Robin17}
{Robin} A.~C., {Bienaym{\'e}} O., {Fern{\'a}ndez-Trincado} J.~G., {Reyl{\'e}}
  C., 2017, \aap, 605, A1

\bibitem[{{Robin} {et~al}\mbox{.}(2003){Robin}, {Reyl{\'e}}, {Derri{\`e}re}, \&
  {Picaud}}]{Robin03}
{Robin} A.~C., {Reyl{\'e}} C., {Derri{\`e}re} S., {Picaud} S., 2003, \aap, 409,
  523

\bibitem[{{Schlafly} \& {Finkbeiner}(2011)}]{Schlafly11}
{Schlafly} E.~F., {Finkbeiner} D.~P., 2011, \apj, 737, 103

\bibitem[{{Schultz} \& {Wiemer}(1975)}]{Schultz75}
{Schultz} G.~V., {Wiemer} W., 1975, \aap, 43, 133

\bibitem[{{Sestito} {et~al}\mbox{.}(2019){Sestito}, {Longeard}, {Martin},
  {Starkenburg}, {Fouesneau}, {Gonz{\'a}lez Hern{\'a}ndez}, {Arentsen},
  {Ibata}, {Aguado}, {Carlberg}, {Jablonka}, {Navarro}, {Tolstoy}, \&
  {Venn}}]{Sestito19}
{Sestito} F. {et~al.}, 2019, \mnras, 484, 2166

\bibitem[{{Sestito} {et~al}\mbox{.}(2023{\natexlab{a}}){Sestito}, {Venn},
  {Arentsen}, {Aguado}, {Kielty}, {Lardo}, {Martin}, {Navarro}, {Starkenburg},
  {Waller}, {Carlberg}, {Fran{\c{c}}ois}, {Gonz{\'a}lez Hern{\'a}ndez},
  {Kordopatis}, {Vitali}, \& {Yuan}}]{Sestito23}
---, 2023{\natexlab{a}}, \mnras, 518, 4557

\bibitem[{{Sestito} {et~al}\mbox{.}(2023{\natexlab{b}}){Sestito}, {Zaremba},
  {Venn}, {D'Aoust}, {Hayes}, {Jensen}, {Navarro}, {Jablonka},
  {Fern{\'a}ndez-Alvar}, {Glover}, {McConnachie}, \&
  {Chen{\'e}}}]{Sestito23Umi}
---, 2023{\natexlab{b}}, arXiv e-prints, arXiv:2301.13214

\bibitem[{{Shapley}(1938)}]{Shapley38}
{Shapley} H., 1938, Harvard College Observatory Bulletin, 908, 1

\bibitem[{{Simon}(2019)}]{Simon19}
{Simon} J.~D., 2019, \araa, 57, 375

\bibitem[{{Simon} {et~al}\mbox{.}(2015){Simon}, {Jacobson}, {Frebel},
  {Thompson}, {Adams}, \& {Shectman}}]{Simon15}
{Simon} J.~D., {Jacobson} H.~R., {Frebel} A., {Thompson} I.~B., {Adams} J.~J.,
  {Shectman} S.~A., 2015, \apj, 802, 93

\bibitem[{{Sk{\'u}lad{\'o}ttir} {et~al}\mbox{.}(2021){Sk{\'u}lad{\'o}ttir},
  {Salvadori}, {Amarsi}, {Tolstoy}, {Irwin}, {Hill}, {Jablonka}, {Battaglia},
  {Starkenburg}, {Massari}, {Helmi}, \& {Posti}}]{Skuladottir21}
{Sk{\'u}lad{\'o}ttir} {\'A}. {et~al.}, 2021, \apjl, 915, L30

\bibitem[{{Sneden}(1973)}]{Sneden73}
{Sneden} C.~A., 1973, PhD thesis, THE UNIVERSITY OF TEXAS AT AUSTIN.

\bibitem[{{Starkenburg} {et~al}\mbox{.}(2013){Starkenburg}, {Hill}, {Tolstoy},
  {Fran{\c{c}}ois}, {Irwin}, {Boschman}, {Venn}, {de Boer}, {Lemasle},
  {Jablonka}, {Battaglia}, {Groot}, \& {Kaper}}]{Starkenburg13}
{Starkenburg} E. {et~al.}, 2013, \aap, 549, A88

\bibitem[{{Starkenburg} {et~al}\mbox{.}(2010){Starkenburg}, {Hill}, {Tolstoy},
  {Gonz{\'a}lez Hern{\'a}ndez}, {Irwin}, {Helmi}, {Battaglia}, {Jablonka},
  {Tafelmeyer}, {Shetrone}, {Venn}, \& {de Boer}}]{Starkenburg10}
---, 2010, \aap, 513, A34

\bibitem[{{Starkenburg} {et~al}\mbox{.}(2017){Starkenburg}, {Martin},
  {Youakim}, {Aguado}, {Allende Prieto}, {Arentsen}, {Bernard}, {Bonifacio},
  {Caffau}, {Carlberg}, {C{\^o}t{\'e}}, {Fouesneau}, {Fran{\c c}ois}, {Franke},
  {Gonz{\'a}lez Hern{\'a}ndez}, {Gwyn}, {Hill}, {Ibata}, {Jablonka},
  {Longeard}, {McConnachie}, {Navarro}, {S{\'a}nchez-Janssen}, {Tolstoy}, \&
  {Venn}}]{Starkenburg17b}
---, 2017, \mnras, 471, 2587

\bibitem[{{Tafelmeyer} {et~al}\mbox{.}(2010){Tafelmeyer}, {Jablonka}, {Hill},
  {Shetrone}, {Tolstoy}, {Irwin}, {Battaglia}, {Helmi}, {Starkenburg}, {Venn},
  {Abel}, {Francois}, {Kaufer}, {North}, {Primas}, \&
  {Szeifert}}]{Tafelmeyer10}
{Tafelmeyer} M. {et~al.}, 2010, \aap, 524, A58

\bibitem[{{Taylor}(2005)}]{Taylor05}
{Taylor} M.~B., 2005, in Astronomical Society of the Pacific Conference Series,
  Vol. 347, Astronomical Data Analysis Software and Systems XIV, {Shopbell} P.,
  {Britton} M., {Ebert} R., eds., p.~29

\bibitem[{{The Dark Energy Survey Collaboration}(2005)}]{DES05}
{The Dark Energy Survey Collaboration}, 2005, arXiv e-prints, astro

\bibitem[{{Tody}(1986)}]{Tody86}
{Tody} D., 1986, in Society of Photo-Optical Instrumentation Engineers (SPIE)
  Conference Series, Vol. 627, Society of Photo-Optical Instrumentation
  Engineers (SPIE), {Crawford} D.~L., ed., p. 733

\bibitem[{{Tody}(1993)}]{Tody93}
---, 1993, in Astronomical Society of the Pacific Conference Series, Vol.~52,
  Astronomical Data Analysis Software and Systems II, {Hanisch} R.~J.,
  {Brissenden} R.~J.~V., {Barnes} J., eds., p. 173

\bibitem[{{Tolstoy} {et~al}\mbox{.}(2006){Tolstoy}, {Hill}, {Irwin}, {Helmi},
  {Battaglia}, {Letarte}, {Venn}, {Jablonka}, {Shetrone}, {Arimoto}, {Abel},
  {Primas}, {Kaufer}, {Szeifert}, {Francois}, \& {Sadakane}}]{Tolstoy06}
{Tolstoy} E. {et~al.}, 2006, The Messenger, 123, 33

\bibitem[{{Tolstoy}, {Hill} \& {Tosi}(2009){Tolstoy}, {Hill}, \&
  {Tosi}}]{Tolstoy09}
{Tolstoy} E., {Hill} V., {Tosi} M., 2009, \araa, 47, 371

\bibitem[{{Tolstoy} {et~al}\mbox{.}(2004){Tolstoy}, {Irwin}, {Helmi},
  {Battaglia}, {Jablonka}, {Hill}, {Venn}, {Shetrone}, {Letarte}, {Cole},
  {Primas}, {Francois}, {Arimoto}, {Sadakane}, {Kaufer}, {Szeifert}, \&
  {Abel}}]{Tolstoy04}
{Tolstoy} E. {et~al.}, 2004, \apjl, 617, L119

\bibitem[{{Tolstoy} {et~al}\mbox{.}(2023){Tolstoy}, {Sk{\'u}lad{\'o}ttir},
  {Battaglia}, {Brown}, {Massari}, {Irwin}, {Starkenburg}, {Salvadori}, {Hill},
  {Jablonka}, {Salaris}, {van Essen}, {Olsthoorn}, {Helmi}, \&
  {Pritchard}}]{Tolstoy23}
---, 2023, arXiv e-prints, arXiv:2304.11980

\bibitem[{{Venn} {et~al}\mbox{.}(2004){Venn}, {Irwin}, {Shetrone}, {Tout},
  {Hill}, \& {Tolstoy}}]{Venn04}
{Venn} K.~A., {Irwin} M., {Shetrone} M.~D., {Tout} C.~A., {Hill} V., {Tolstoy}
  E., 2004, \aj, 128, 1177

\bibitem[{{Walker} \& {Pe{\~n}arrubia}(2011)}]{Walker11}
{Walker} M.~G., {Pe{\~n}arrubia} J., 2011, \apj, 742, 20

\bibitem[{{Waller} {et~al}\mbox{.}(2023){Waller}, {Venn}, {Sestito}, {Jensen},
  {Kielty}, {Borukhovetskaya}, {Hayes}, {McConnachie}, \& {Navarro}}]{Waller23}
{Waller} F. {et~al.}, 2023, \mnras, 519, 1349

\bibitem[{{Wenger} {et~al}\mbox{.}(2000){Wenger}, {Ochsenbein}, {Egret},
  {Dubois}, {Bonnarel}, {Borde}, {Genova}, {Jasniewicz}, {Lalo{\"e}},
  {Lesteven}, \& {Monier}}]{Wenger00}
{Wenger} M. {et~al.}, 2000, \aaps, 143, 9

\bibitem[{{Westfall} {et~al}\mbox{.}(2006){Westfall}, {Majewski}, {Ostheimer},
  {Frinchaboy}, {Kunkel}, {Patterson}, \& {Link}}]{Westfall06}
{Westfall} K.~B., {Majewski} S.~R., {Ostheimer} J.~C., {Frinchaboy} P.~M.,
  {Kunkel} W.~E., {Patterson} R.~J., {Link} R., 2006, \aj, 131, 375

\bibitem[{{Wheeler} {et~al}\mbox{.}(2019){Wheeler}, {Hopkins}, {Pace},
  {Garrison-Kimmel}, {Boylan-Kolchin}, {Wetzel}, {Bullock}, {Kere{\v{s}}},
  {Faucher-Gigu{\`e}re}, \& {Quataert}}]{Wheeler19}
{Wheeler} C. {et~al.}, 2019, \mnras, 490, 4447

\bibitem[{{White} {et~al}\mbox{.}(1987){White}, {Davis}, {Efstathiou}, \&
  {Frenk}}]{White87}
{White} S. D.~M., {Davis} M., {Efstathiou} G., {Frenk} C.~S., 1987, \nat, 330,
  451

\bibitem[{{White} \& {Rees}(1978)}]{White78}
{White} S.~D.~M., {Rees} M.~J., 1978, \mnras, 183, 341

\bibitem[{{Yang} {et~al}\mbox{.}(2022){Yang}, {Hammer}, {Jiao}, \&
  {Pawlowski}}]{Yang22}
{Yang} Y., {Hammer} F., {Jiao} Y., {Pawlowski} M.~S., 2022, \mnras, 512, 4171

\bibitem[{{York} {et~al}\mbox{.}(2000){York}, {Adelman}, {Anderson},
  {Anderson}, {Annis}, {Bahcall}, {Bakken}, {Barkhouser}, {Bastian}, {Berman},
  {Boroski}, {Bracker}, {Briegel}, {Briggs}, {Brinkmann}, {Brunner}, {Burles},
  {Carey}, {Carr}, {Castander}, {Chen}, {Colestock}, {Connolly}, {Crocker},
  {Csabai}, {Czarapata}, {Davis}, {Doi}, {Dombeck}, {Eisenstein}, {Ellman},
  {Elms}, {Evans}, {Fan}, {Federwitz}, {Fiscelli}, {Friedman}, {Frieman},
  {Fukugita}, {Gillespie}, {Gunn}, {Gurbani}, {de Haas}, {Haldeman}, {Harris},
  {Hayes}, {Heckman}, {Hennessy}, {Hindsley}, {Holm}, {Holmgren}, {Huang},
  {Hull}, {Husby}, {Ichikawa}, {Ichikawa}, {Ivezi{\'c}}, {Kent}, {Kim},
  {Kinney}, {Klaene}, {Kleinman}, {Kleinman}, {Knapp}, {Korienek}, {Kron},
  {Kunszt}, {Lamb}, {Lee}, {Leger}, {Limmongkol}, {Lindenmeyer}, {Long},
  {Loomis}, {Loveday}, {Lucinio}, {Lupton}, {MacKinnon}, {Mannery}, {Mantsch},
  {Margon}, {McGehee}, {McKay}, {Meiksin}, {Merelli}, {Monet}, {Munn},
  {Narayanan}, {Nash}, {Neilsen}, {Neswold}, {Newberg}, {Nichol}, {Nicinski},
  {Nonino}, {Okada}, {Okamura}, {Ostriker}, {Owen}, {Pauls}, {Peoples},
  {Peterson}, {Petravick}, {Pier}, {Pope}, {Pordes}, {Prosapio},
  {Rechenmacher}, {Quinn}, {Richards}, {Richmond}, {Rivetta}, {Rockosi},
  {Ruthmansdorfer}, {Sand ford}, {Schlegel}, {Schneider}, {Sekiguchi},
  {Sergey}, {Shimasaku}, {Siegmund}, {Smee}, {Smith}, {Snedden}, {Stone},
  {Stoughton}, {Strauss}, {Stubbs}, {SubbaRao}, {Szalay}, {Szapudi}, {Szokoly},
  {Thakar}, {Tremonti}, {Tucker}, {Uomoto}, {Vanden Berk}, {Vogeley},
  {Waddell}, {Wang}, {Watanabe}, {Weinberg}, {Yanny}, {Yasuda}, \& {SDSS
  Collaboration}}]{York00}
{York} D.~G. {et~al.}, 2000, \aj, 120, 1579

\bibitem[{{Zhang} {et~al}\mbox{.}(2012){Zhang}, {Hunter}, {Elmegreen}, {Gao},
  \& {Schruba}}]{Zhang12}
{Zhang} H.-X., {Hunter} D.~A., {Elmegreen} B.~G., {Gao} Y., {Schruba} A., 2012,
  \aj, 143, 47

\end{thebibliography}

\bsp	
\label{lastpage}
\end{document}